\documentclass{article}

\usepackage{arxiv}

\usepackage[utf8]{inputenc} 
\usepackage[T1]{fontenc}    
\usepackage{hyperref}       
\usepackage{url}            
\usepackage{booktabs}       
\usepackage{amsfonts}       
\usepackage{nicefrac}       
\usepackage{microtype}      
\usepackage{lipsum}
\usepackage{graphicx,verbatim,float,subfigure,graphicx,color,siunitx}
\usepackage{amsmath}
\graphicspath{ {./images/} }

\usepackage{amssymb}
\usepackage{svg}
\usepackage{multirow}  
\usepackage[export]{adjustbox}
\usepackage{soul}
\usepackage{textcomp}
\usepackage{multirow}
\usepackage{enumitem}
\usepackage{amsfonts}
\usepackage{bbm}
\usepackage{amsthm,lineno}

\newlist{Properties}{enumerate}{2}
\setlist[Properties]{label=Property \arabic*., font=\textbf, itemindent=*}

\newtheorem{Proposition}{Proposition}
\newtheorem{propositionproof}{Proof of Proposition}

\title{Physics-Constrained Neural Network for Design and Feature-Based Optimization of Weave Architectures}

\author{
 Haotian Feng \\
  Dept. of Mechanical Engineering\\
  University of Wisconsin-Madison \\
  Madison, WI 53706 \\
  \And
 Sabarinathan P Subramaniyan \\
  Dept. of Mechanical Engineering \\
  University of Wisconsin-Madison \\
  Madison, WI 53706 
  \And
Hridyesh R. Tewani \\
  Dept. of Civil \& Env. Engineering \\
  University of Wisconsin-Madison \\
  Madison, WI 53706 
  \And
  Pavana Prabhakar* \\
  Dept. of Mechanical Engineering \\
  Dept. of Civil \& Env. Engineering \\
  University of Wisconsin-Madison \\
  Madison, WI 53706 \\
  \texttt{pavana.prabhakar@wisc.edu} \\
}

\begin{document}
\maketitle

\newcommand{\SPS}[1]{\textcolor{red}{\bf{Sabari: #1}}}

\begin{abstract}

Woven fabrics play an essential role in everyday textiles for clothing/sportswear, water filtration, and retaining walls, to reinforcements in stiff composites for lightweight structures like aerospace, sporting, automotive, and marine industries. Several possible combinations of weave patterns and material choices, which comprise weave architecture, present a challenging question about how they could influence the physical and mechanical properties of woven fabrics and reinforced structures. In this paper, we present a novel Physics-Constrained Neural Network (PCNN) to predict the mechanical properties like the modulus of weave architectures and the inverse problem of predicting pattern/material sequence for a design/target modulus value. The inverse problem is particularly challenging as it usually requires many iterations to find the appropriate architecture using traditional optimization approaches. We show that the proposed PCNN can effectively predict weave architecture for the desired modulus with higher accuracy than several baseline models considered. We present a feature-based optimization strategy to improve the predictions using features in the Grey Level Co-occurrence Matrix (GLCM) space. We combine PCNN with this feature-based optimization to discover near-optimal weave architectures to facilitate the initial design of weave architecture. The proposed frameworks will primarily enable the woven composite analysis and optimization process, and be a starting point to introduce Knowledge-guided Neural Networks into the complex structural analysis.


\end{abstract}


\section{Introduction}\label{intro}
Woven fabric, a textile material, is formed by weaving or interlacing warp and weft fiber bundles in the orthogonal directions. Woven fabric has a wide range of applications, from everyday textiles for clothing and fashion to reinforcements in stiff composites for lightweight structures like aerospace, sporting, automotive, and marine industries \cite{mouritz1999review,kelkar2006structural,carey2017introduction,gao2021textile}. Possible combinations of weave patterns and choices of materials for the warp/weft fiber bundles present a promising yet challenging question about how they could influence corresponding physical and mechanical properties. To that end, we present a novel \textbf{Physics-Constrained Neural Network (PCNN)} to predict the mechanical properties like the modulus of weave architectures (weave pattern, weave material sequence) and the inverse problem of predicting pattern/material sequence for a design/target modulus value. Though these frameworks can be applied to any woven fabric, we consider \textbf{woven composites} as our case study to develop and demonstrate their phenomenal advantage.

Woven composites are stiff composites comprising woven fabrics with high-strength fibers like Carbon, glass, Aramid, etc., reinforced within polymers. These composites have drawn significant interest in recent years due to their tunable mechanical properties, high strength-to-weight ratio, high production rate, and structural durability\cite{long2005design}. Due to these significant advantages, textile composites have been used in the aerospace, sporting, automotive, and marine industries. To better understand woven composites, many researchers have been focusing on exploring the mechanical properties of woven composites. Research to find woven composite's mechanical properties largely relies on analytical representation or numerical analysis like Finite Element Analysis (FEA). Research to understand woven composite first focuses on the analytical representation of the woven composite model. Naik et al.\cite{naik1992prediction} proposed a shape function to define the woven fabric geometry by considering the actual strand cross-section geometry, the possible gap between adjacent strands, and the undulation and continuity of strands along the warp and weft directions. Jiang et al.\cite{jiang2006investigation} presented a three-dimensional representative volume-element model to study the micromechanical behavior of woven fabric composites; the model displayed a good agreement with the published experimental. Moreover, the relationship between geometric parameters and the macromechanical behavior of the composites could be obtained from the proposed model. Khan et al.\cite{khan2017finite} proposed a simplified mathematical micromechanics model for calculating the mechanical properties of plain weave composites using FEA. The proposed model considered geometry close to the actual fabric by utilizing geometric parameters like yarn undulations and interactions between warp and weft tows. Although the analytical approach is computationally efficient, it cannot accurately represent the model's complexity and mechanical responses. Thus, several researchers focus on utilizing FEA to analyze the woven composite models numerically. Ishikawa et al.\cite{ishikawa1983one} conduct the one-dimensional micromechanical analysis on the woven composite to derive the upper and lower bounds of stiffness and compliance constants. The result is further validated with 2D FEA. Whitcomb et al.\cite{whitcomb1991three,whitcomb1994macro,whitcomb1995boundary} utilize FEA to analyze the three-dimensional stress of plain woven composite and the boundary effect of woven composites. Gowayed et al.\cite{gowayed2013types} presented different types of fiber and fiber arrangements in fiber-reinforced polymer woven fabrics. The impact of fiber assembling into yarns and fabrics is also discussed in the paper. Dong et al.\cite{dong2016experimental} utilize experimental and Finite Element analysis to find the plain weave composite's thermal conductivity and further compare the conductive behavior with unidirectional lamina. These methods have shown the power of FEA in analyzing woven composite models by including much more geometric complexity than the analytical approach. However, using FEA to explore the mechanical properties of the woven composite is time-consuming as each woven model needs to be solved numerically. The time consumption is even more for optimizing the weave patterns for specified properties. 

The emergence of Machine Learning (ML) methods research largely facilitates understanding composite materials and predicting the corresponding mechanical properties. Among existing ML algorithms, Deep Convolutional Neural Network\cite{krizhevsky2012imagenet} (DCNN) and Generative Adversarial Network\cite{creswell2018generative} (GAN) are the most widely used. DCNN is a class of deep neural networks consisting of several convolutional, pooling, and fully connected layers. DCNN has been widely used in different fields, including image classification\cite{krizhevsky2012imagenet}, recommender system\cite{ying2018graph}, image segmentation\cite{zhang2015deep}, and natural language processing\cite{conneau2016very}. GAN is developed similarly to game theory, where Nash equilibrium is reached when the model converges. There is a generator and a discriminator Network in GAN. GAN has been used in different fields, including unsupervised learning\cite{schlegl2017unsupervised,bousmalis2017unsupervised},  semi-supervised learning\cite{souly2017semi}, fully supervised learning\cite{zhao2016energy}, and reinforcement learning\cite{yu2017seqgan}. Regarding ML's application in composite material analysis, Wei et al.\cite{wei2018predicting} demonstrate that machine learning methods like support vector regression, Gaussian process regression, and convolutional neural network (CNN) are useful tools to predict the effective thermal conductivities of composite materials and porous media. Chen et al.\cite{chen2019machine} give an overview of how different Machine Learning algorithms can accelerate composite material research, including several different regression models, Neural networks (especially CNN), and the Gaussian process. Feng et al.\cite{feng2021difference} propose a Deep Learning method to predict composite micromechanical models' stress distribution contours using a Difference-based Neural Network, where the Neural Network focuses on predicting the differences to a reference sample. Bang et al.\cite{bang2020defect} propose a framework to identify the defects within composite material by integrating thermo-graphic images of composite with deep learning. Liu et al.\cite{liu2019initial} propose a new failure criterion for fiber tows in woven composite by combining mechanics of structure genome and a deep neural network model. Nardi et al.\cite{nardi2021design} utilize the Artificial Neural Network to predict the thermoforming process of thermoplastic composites. The authors focus on the glass fiber-reinforced polyetherimide woven composite and discuss the essential features needed for accurate predictions of the temperature fields over the thermoforming process. The authors further discuss the potentiality of using Machine Learning to determine the optimal range of the process parameters. Sepasdar et al.\cite{sepasdar2021data} propose the modified U-Net network to predict the damage and failure in microstructure-dependent composite materials. Gu et al.\cite{gu2018bioinspired} use ML to analyze the strength and toughness of 2D checkerboard models for 2D printed bi-material composites. The authors used a single-layer convolutional neural network with two binary classifiers. Further, Abueidda et al.\cite{abueidda2019prediction} also focus on a 2D checkerboard model and utilized a genetic algorithm optimizer to optimize a checkerboard composite pattern to obtain a model with maximum strength and toughness based on different volume fractions. This research has proven the potentiality of accelerating woven composite design and analysis with ML. 

Although insightful, these above frameworks are limited to predicting material properties for a given pattern or optimizing through heuristic searching, which is relatively easy to handle. On the contrary, the ability to solve the inverse design problem, which predicts patterns for target mechanical properties, can be more challenging and beneficial. Within woven composites, it could save a massive amount of time otherwise invested in testing weave design iterations. Feng et al.\cite{feng2021deep} considered 2D woven composite with a single material and proposed the GAN-based framework for the inverse design problem. The research has shown the potentiality of utilizing Neural networks with a relatively decent error rate of around 7\%. Similarly, Chen et al.\cite{chen2020generative} consider the inverse design problem of the checkerboard composite model using generative inverse design networks called GIDN. GIDN consists of a predictor and a designer, like the idea of GAN. The predictor is first trained with training data, then trained weights in the predictor are directly assigned to the designer as non-trainable parameters. The designer further provides an optimized design from the initial Gaussian distributed design. GIDN has outperformed conventional gradient-based topology optimization and gradient-free algorithms for a stiff-soft bi-material composite model. This method brings promising ideas to optimize the composite material, while this GAN-based approach does not build the connection between the mechanical properties of composite material to its geometry. Also, Neural Network-based optimization is hard to be understood in the physical space. 

Thus, this paper aims to solve two problems related to 2D woven composite, whose pattern can also be represented as a checkerboard model: (1) How can we build a bi-directional bridge between woven composite architecture and its mechanical properties? (2) How to optimize the woven composite's mechanical properties using 'physically meaningful' features, so we can optimize the woven composite properties by directly manipulating physical and geometric parameters?

\section{Motivation and Overview}
This section presents an overview of the overall targets of the research presented in this paper and the general Machine Learning approaches that we use for woven composite prediction and optimization.

\subsection{Motivation and Research Tasks}

As mentioned before, in this work, we focus on solving two problems related to understanding the mechanical properties of woven composite and optimizing the woven architecture to achieve improved overall in-plane modulus. Woven composite architecture could be determined by different combinations of weave patterns and material sequences. For example, a 6-by-6 woven composite model will have $2^{36}$ different patterns and $2*n^6$ different material sequences, where $n$ is the number of materials to choose from. Thus, it is essential to efficiently and accurately obtain the mechanical properties of different woven composite architectures to determine the optimal architecture suitable for the problem of interest. Besides understanding the mechanical behavior of woven composites, optimization is also critical to minimizing the structure's stresses, weight, or compliance for a given amount of material and boundary conditions. Through optimization, we want to determine the most advantageous structure or material distribution that results in the highest mechanical properties for the design requirement.

\begin{figure}[h!]
\centering
	\includegraphics[width=0.95\textwidth]{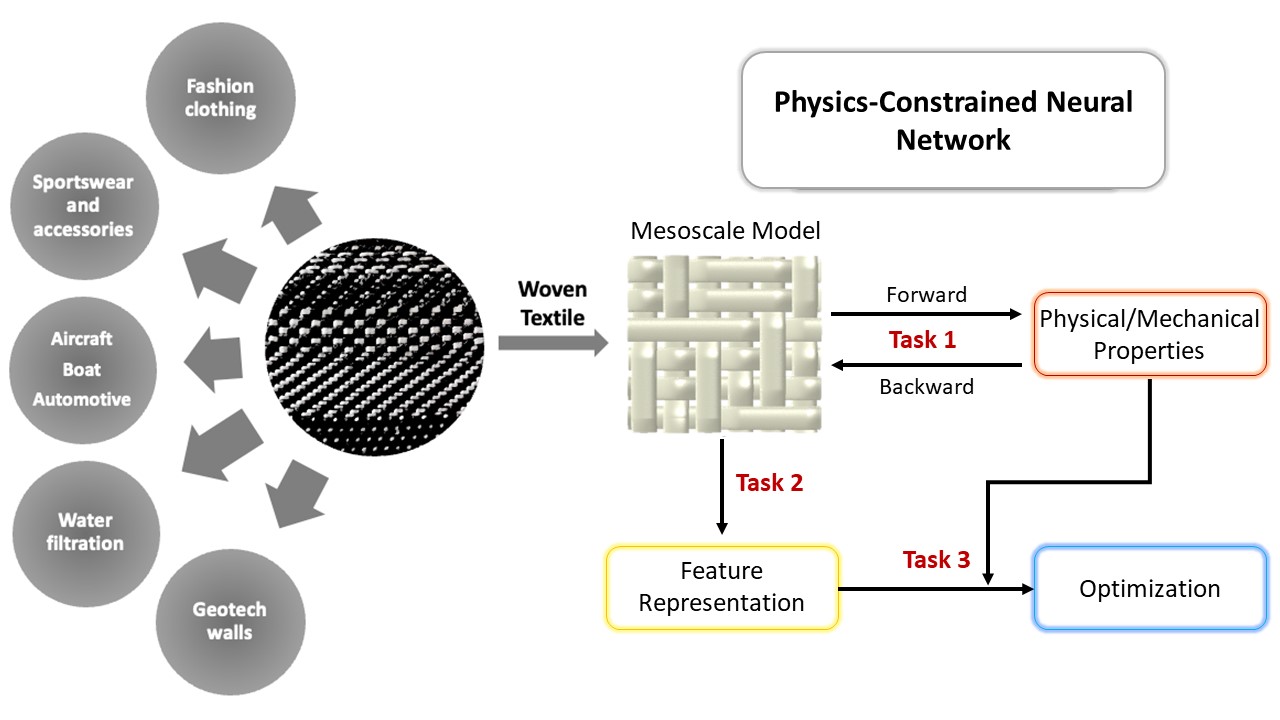}
\caption{Overview of three Machine Learning tasks: (1) \textbf{Task 1} builds the bridge between woven composite and its physical/mechanical properties. Task 1 is split into Forward Direction Prediction (FDP) and Backward Direction Prediction (BDP) problems. BDP is further split into BDPa and BDPb, depending on predicting the weave pattern or material sequence from a given target value of in-plane moduli. (2) \textbf{Task 2} proposes the feature-based statistical representation of woven composite for weave pattern and material sequence and describes the relationship between extracted features and corresponding mechanical/physical properties through statistical analysis. (3) \textbf{Task 3} proposes the optimization strategy on the woven composite to achieve improved physical/mechanical properties (like higher strength) and discovers the near-optimal woven composite design using the methods developed in Tasks 1 and 2. (Here, 'near-optimal' refers to an improved  woven composite design but might not guarantee to be the global optimal design.)
}
\label{img:paper_overview}
\end{figure}

In this paper, we consider two different woven composite models: single-material and bi-material woven composite. Single-material woven composites consist of yarns made of one material for the whole model, whereas bi-material woven composites have different yarns made of two materials. Specifically, we will consider three tasks (an overview is represented in Figure~\ref{img:paper_overview}) as follows: 
\begin{enumerate}
    \item \textbf{Task 1}: Establish the connection between woven composite architecture (pattern + material) and corresponding in-plane moduli. We will focus on the following tasks: (1) Forward Direction Prediction (FDP): predicting from woven composite architecture to the corresponding modulus. (2) Backward Direction Prediction (BDP): predicting from woven modulus to its architecture. 
    We decouple the BDP problem into two sub-problems: prediction from weave in-plane modulus and material sequence to its pattern (named as BDPa) and prediction from weave in-plane modulus and pattern to its material sequence (named as BDPb).
    \item \textbf{Task 2}: Propose a feature-based statistical representation of the woven composite. Specifically, we propose representing the weave pattern using the Gray Level Co-occurrence Matrix (GLCM). We prove the uniqueness of GLCM statistical features from a binary matrix and how each statistical feature is related to the weave pattern feature in the physical space. We further represent the weave material sequence with features from the physical space. Later, we conduct statistical analysis to understand how each feature is correlated with the overall moduli ($E_{all}=E_1+E_2+G_{12}$) of woven composites.
    \item \textbf{Task 3}: Propose the feature-based statistical optimization strategy to find the woven composite with the highest overall moduli and discover the near-optimal woven composite design using the methods developed in Task 1 and Task 2. From the statistical analysis, we can determine whether each statistical feature is positively or negatively correlated with the woven composite's overall in-plane modulus and further optimize the choice of weave pattern and material sequence based on such correlation relationship.
\end{enumerate}

For \textbf{Task 1}, to solve the FDP problem, we utilize Deep Convolutional Neural Network (DCNN) to extract high-level features from the woven composite model and predict the in-plane modulus from its architecture. The BDP problems are more challenging than the FDP problems since the in-plane modulus can be sensitive to weave patterns and material sequences. Incorrect prediction at a single position in the pattern or material sequence could significantly change the in-plane modulus. Moreover, we will show that woven composite with different patterns could have similar in-plane modulus. Such similarity forms one-to-many mapping for BDP problems. So, a purely data-driven Neural Network makes it hard to achieve high accuracy in BDP problems. To constrain the predictions for BDP problems, we combine the idea of the existing Physics-Informed Neural Network (PINN)\cite{mao2020physics} and transfer learning\cite{pan2009survey,weiss2016survey}, and then propose the \textbf{Physics-Constrained Neural Network (PCNN)}, which embeds the existing physics knowledge into the Neural Network to constrain the prediction. Specifically, PCNN will first utilize a similar structure as deep convolutional Autoencoder\cite{badrinarayanan2017segnet} to extract high-level features from the input data and make predictions based on these extracted features. Then, the PCNN will simultaneously embed our physics knowledge in the prediction layer and contribute certain losses to the loss function. Here, the physics knowledge refers to the relationship between woven composite architecture (pattern + material sequence) and its corresponding modulus, which comes from the trained DCNN in the FDP problem. We further validated that our proposed PCNN could enhance prediction accuracy compared to many widely used machine learning frameworks for BDPa and BDPb problems.

For \textbf{Task 2}, we consider the feature-based statistical representation of weave patterns, which can be represented as a checkerboard model and treated as a type of texture. Then, we extract texture features from the weave pattern. Texture features describe the spatial distribution of pixels (cells), which reflect objects' roughness, smoothness, granularity, and randomness. Common texture feature extraction methods include statistical, structural, and spectral methods. This paper utilizes the statistical method and proposes the GLCM feature-based optimization strategy. GLCM, referring to Gray-Level Co-Occurrence Matrix, is a statistical method of examining texture that considers the spatial relationship of pixels\cite{sebastian2012gray}. The GLCM features characterize an image's texture by calculating how often pairs of pixels with specific values and in a specified spatial relationship occur in an image and then extracting statistical measures from the matrix. Since GLCM can measure the texture roughness, coarseness, and other properties in one calculation, it has been the primary method to describe texture-related methods in the field of medical sciences (CT scans, MRI)\cite{zulpe2012glcm,singh2012classification}, landscape analysis\cite{hall2017glcm} and image-based defect detection\cite{raheja2013fabric}. In this paper, specifically, we use Haralick texture features\cite{haralick1973textural}.
Furthermore, we represent the weave material sequence using the statistical features in the physical space. Two vectors can represent the material sequence, and each vector describes the material sequence for weft and warp yarns. We consider statistical features directly from the material sequence vector, including mean, median, and standard deviation. 

For \textbf{Task 3}, utilizing the statistical features extracted from Task 2, we describe the correlation relationship between extracted GLCM-based Haralick features from weave patterns and the corresponding in-plane modulus of woven composites through statistical analysis to guide weave pattern optimization. Similarly, we determine how each statistical feature is correlated with the in-plane modulus for weave material sequence optimization and determine the optimal material sequence from statistical analysis. Finally, the statistical models based on weave pattern and material sequence can be combined to optimize a given woven architecture, which can be further combined with PCNN to discover near-optimal woven composite architecture at the initial design stage.

\subsection{Overview of Proposed Machine Learning Framework}

\begin{figure}[h!]
\centering
	\includegraphics[width=0.99\textwidth]{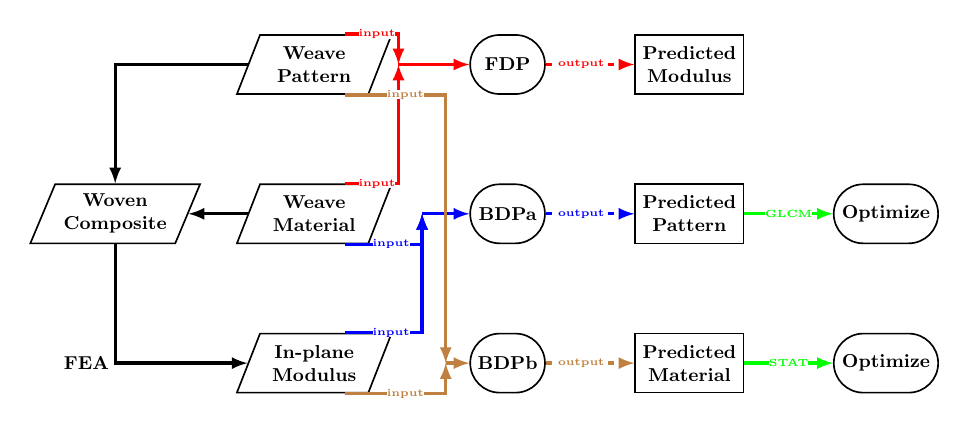}
\caption{Overview of the proposed Machine Learning framework: (1) Black arrows represent the FEA process, (2) Red arrows represent the FDP problem, (3) Blue and Brown arrows represent two BDP problems: BDPa and BDPb. Trapezium blocks are inputs to the Machine Learning framework. Initially, we have weave patterns and material sequences; then, these models are brought into Finite Element solver ABAQUS\cite{ABAQUS} to find the corresponding in-plane modulus. Circular blocks represent different Machine Learning tasks: FDP, BDPa, and BDPb. Square blocks are the predictions for different Machine Learning tasks. Solid lines before circular blocks represent inputs for the Machine Learning framework, and dashed lines represent outputs. The solid green line and rounded corner blocks are the optimization modules. We introduce feature-based optimization for pattern and materials sequence, using GLCM and physical space statistical features.}
\label{img:overview}
\end{figure}

In Figure~\ref{img:overview}, we present the proposed machine-learning framework for the two tasks considered in this paper. First, weave patterns and materials are picked to define each woven composite model uniquely. Then, we can calculate the corresponding in-plane modulus through FEA by applying boundary conditions. After obtaining the weave pattern, material sequence, and corresponding in-plane modulus, we can start the Machine Learning process: (1) For FDP, we design a deep convolutional neural network that takes the weave pattern and material sequence as inputs and outputs the in-plane modulus. (2) For BDPa, we design the \textbf{Physics-Constraint Neural Network (PCNN)} that takes in-plane modulus and material sequence as inputs and predicts the pattern that matches the target in-plane modulus. For BDPb, we design another Physics-constraint Neural Network similar to BDPa, which takes the pattern and in-plane modulus as inputs instead and predicts the possible material sequence that matches the target in-plane modulus.

In this paper, we consider single-material and bi-material woven composites. As a constant vector can represent the material sequence for the single-material woven composite, it will not serve as input to train the Machine Learning framework. On the other hand, for bi-material woven composite, weave pattern, material sequence, and corresponding in-plane modulus will be input to train the Machine Learning framework.

Throughout the paper, the machine learning framework is implemented in TensorFlow 2.5.0 and trained on NVIDIA GeForce RTX 2080 SUPER with 3072 CUDA cores and 1815 MHz frequency. We provide access to our implemented Machine Learning code on our GitHub page, as mentioned in the "Data Availability" section at the end of this paper. The GitHub page provides implementations of our proposed Neural Networks, our baseline models for comparison purposes, and the training data used in this paper.

\section{Finite Element Method for Training Data Generation}
As introduced in Section~\ref{intro}, woven composites are formed by inter-laced yarns impregnated with a resin matrix. The woven composite's effective mechanical property depends on the considered material's property, the cross-sectional geometry of yarn, and the weave pattern. 
In this paper, we utilize FEA to determine the in-plane modulus of woven composite ($E_1$, $E_2$, $G_{12}$) based on different combinations of weave patterns and material sequences. We performed FEA on a repeated unit cell (RUC) to understand the influence of weave patterns on the composite's in-plane effective properties. 
We modeled the weave using TexGen \cite{Lin2011}, where all the geometrical input parameters are listed in Table~\ref{tab:geometric_parameter}. 

\begin{table}[h!]
\centering
\caption{Geometrical parameters for finite element modeling}
\begin{tabular}{cccccc}
\hline
Length L & Width W & Height H & Yarn spacing & Yarn height & Yarn width \\ \hline
6mm             & 6mm            & 0.44mm           & 1mm                 & 0.2mm              & 0.8mm             \\ \hline
\end{tabular}
\label{tab:geometric_parameter}
\end{table}


The FEA was divided into two stages: First, we analyzed the influence of the weave pattern for single-material woven composites. Second, the same process is extended to bi-material woven composites with two different yarn materials, Carbon and Kevlar. The homogenized mechanical properties of the fiber bundles embedded in the polymer matrix are shown in Table~\ref{tab:yarn_parameter}, which were calculated using Chamis micro-mechanical model\cite{Chamis1989}. We assumed the volume fraction of fiber to be 76\% in this paper. Initially, the TexGen python scripting generates 9000 random weave patterns with carbon fiber yarns in woven composites. Later, another 9000 random weave patterns with random hybrid carbon-kevlar woven composites were generated. Each geometric model is exported as an input file with linear tetrahedron elements and periodic boundary conditions. In this paper, edge forces are applied in different directions. The corresponding displacement values were extracted from the applied tensile (shear) loading to evaluate the effective in-plane mechanical properties. A detailed explanation of boundary condition implementation can be found in Li et al.\cite{Li2004}. After preprocessing, we imported the input file into ABAQUS to determine the effective in-plane mechanical properties from the stress and displacement field.

\begin{table}[h!]
\centering
\caption{Homogenized material properties of fiber yarn embedded in polymer matrix}
\begin{tabular}{cccccccccc}
\hline
            & $E_1$ (GPa) & $E_2$  (GPa) & $E_3$ (GPa) & $G_{12}$  (GPa) & $G_{13}$ (GPa) & $G_{23}$ (GPa) & $\nu_{12}$ & $\nu_{13}$ & $\nu_{23}$ \\ \hline
Carbon yarn & 183.1      & 9.67        & 9.67       & 5.66         & 5.66        & 3.37        & 0.23   & 0.23   & 0.43   \\
Kevlar yarn & 116.03     & 3.96        & 3.96       & 2.45         & 2.45        & 1.69        & 0.35   & 0.35   & 0.45   \\ \hline
\end{tabular}
\label{tab:yarn_parameter}
\end{table}

\section{Machine Learning Model Inputs}\label{sec:ML_inputs}
We will establish the bridge between weave pattern, material sequence, and in-plane modulus ($E_1$, $E_2$, and $G_{12}$) through Deep Neural Networks. To transform these input data to fit the Neural Network training, we conduct data pre-processing to convert weave patterns and material sequences into matrices and vectors, respectively. 

\subsection{Weave Pattern Representation}\label{sec:weave_pattern}
\begin{figure}[h!]
\centering
	\includegraphics[width=0.9\textwidth]{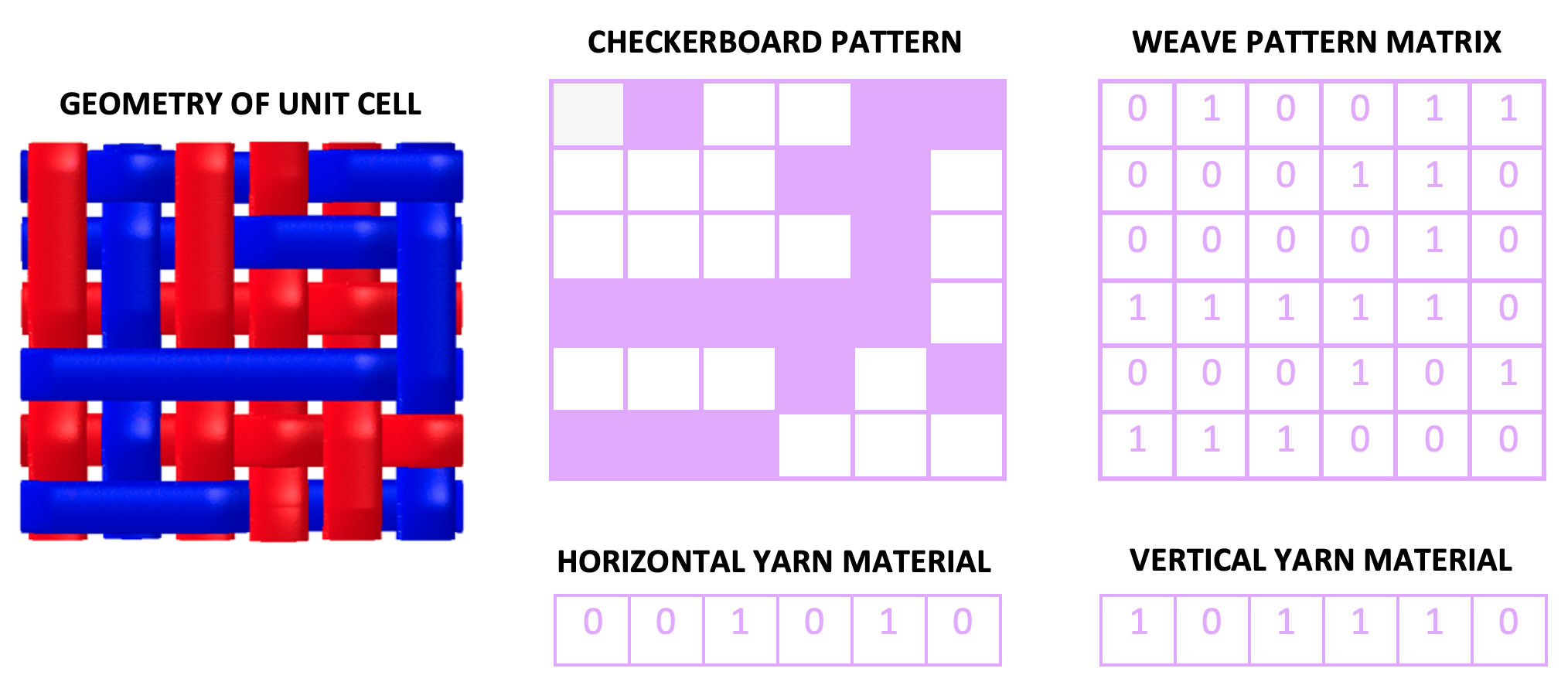}
\caption{Weave pattern and material sequence representation for bi-material woven composite}
\label{img:woven_pattern}
\end{figure}

The yarn placed along the x-axis is called weft, whereas the yarn along the y-axis is called warp. A checkerboard model represents each weave pattern as a matrix with `0' or `1' binary values, where `1' means warp lies below the weft and `0' means warp lies above the weft. We denote Carbon yarn as material `0' and Kevlar yarn as `1' for material sequence. A weave pattern and material sequence  for a bi-material woven composite are shown in Figure~\ref{img:woven_pattern}. This paper considers a woven composite unit cell size of 6-by-6, although others could consider larger unit sizes. So, each model is formed by weaving together six warp and six weft yarns, and a 6-by-6 binary value matrix can represent each pattern.

Previously, researchers have studied woven composites and depicted them in a physical manner. Ishikawa and Chou \cite{ishikawa1982} proposed a "mosaic model" to analyze the mechanical performance of woven composites using analytical approaches. They used a geometrical factor ${n_g}$ to describe the number of warps interlaced with a single weft yarn. They also showed that weave patterns with smaller ${n_g}$ displayed inferior properties due to a higher number of undulations. Further, the bridging model was also proposed to highlight the effect of higher ${n_g}$ on the mechanical performance of woven composites. It was shown that weaves with higher ${n_g}$ values will contain straight yarns in the vicinity of the undulated region. This "quasi-crossply" area would have higher local moduli values and serve as a bridge between the neighboring undulated regions, resulting in higher in-plane moduli for the entire weave structure. We have provided a detailed description of the effect of physical factors on the mechanical performance of woven composites in Section~\ref{app:WeavePar}. This work will use these parameters to justify the optimized weave patterns obtained from the GLCM optimization module.  

\subsection{Weave Material Representation}
Since the material sequence for single-material woven composites will not serve as input to the Neural Network, only bi-material woven composites need the proper representation of their material sequences. As mentioned in Section~\ref{sec:weave_pattern}, the woven composite is formed with six warp and six weft yarns, so the material sequence can be represented as two 6-by-1 binary vectors: the first vector represents warp materials, and the second vector represents weft materials.  

\subsection{In-plane Modulus from FEA Results}
\begin{figure}[h!]
\centering
\subfigure[]{
  \includegraphics[width=0.3\textwidth]{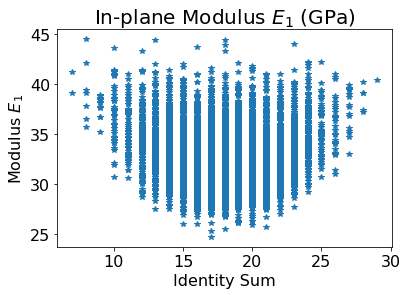}
}
\centering
\subfigure[]{
  \includegraphics[width=0.3\textwidth]{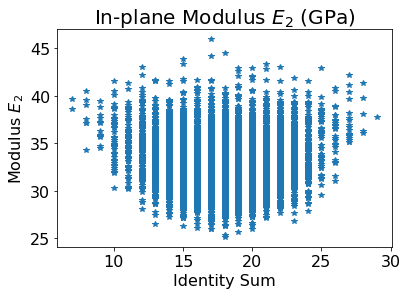}
}
\centering
\subfigure[]{
  \includegraphics[width=0.3\textwidth]{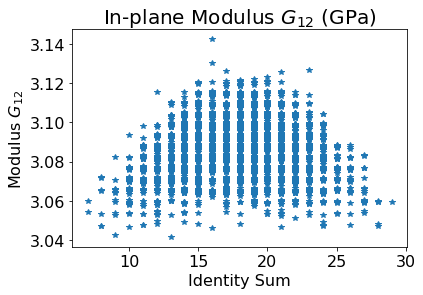}
}
\caption{In-plane moduli distributions for 9000 single material woven composites}
\label{img:single-mat_modulus}
\end{figure}

\begin{figure}[h!]
\centering
\subfigure[]{
  \includegraphics[width=0.3\textwidth]{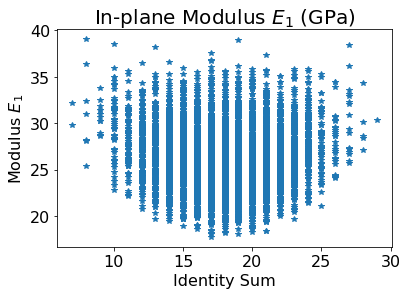}
}
\centering
\subfigure[]{
  \includegraphics[width=0.3\textwidth]{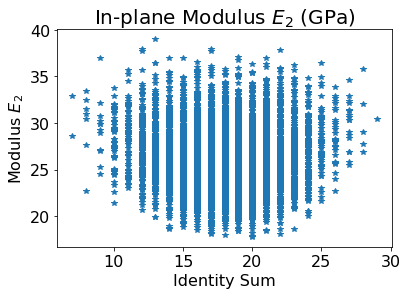}
}
\centering
\subfigure[]{
  \includegraphics[width=0.3\textwidth]{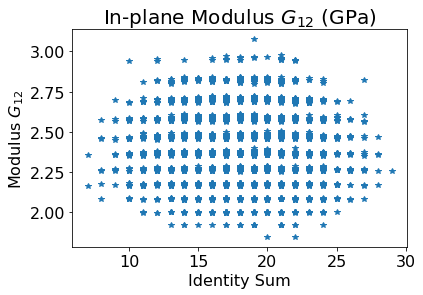}
}
\caption{In-plane moduli distributions for 9000 bi-material woven composite}
\label{img:bi-mat-modulus}
\end{figure}

From the outputs of  9000 single-material and 9000 bi-material woven composite models, we obtain the distributions of different in-plane modulus ($E_1$, $E_2$, $G_{12}$). Here we define the Identity Sum (IS) of woven composite to be: $IS = \sum_{i=1}^{n_1}\sum_{j=1}^{n_2}\mathbbm{1}_{[W_{ij}=1]}$, where $W$ is the weave pattern matrix, $n_1=n_2=6$ as the pattern matrix is 6-by-6. IS of a model represents the total number of `1' regions within the matrix. Figure~\ref{img:single-mat_modulus} and Figure~\ref{img:bi-mat-modulus} show the distribution of different in-plane modulus with respect to identity sum for single material and bi-material woven composites. Comparing these two figures, we discover that: (1) single-material and bi-material woven composites have similar distribution for tensile moduli, $E_1$ and $E_2$; (2) in-plane shear modulus $G_{12}$ distribution for single-material woven composites is more concentrated compared to bi-material woven composites. 

\subsection{Many-To-One Mapping} \label{sec:many-to-one}
From Figure~\ref{img:single-mat_modulus} and Figure~\ref{img:bi-mat-modulus}, we can also observe that woven composites with the same IS could have a completely different in-plane modulus. Moreover, our FEA outputs show that different woven composite patterns could have similar in-plane moduli values. For example, as shown in Appendix~\ref{App:Woven_Hist} - Figure~\ref{img:same_E1}, although the patterns look entirely different, both single material woven composite models have the same modulus in the vertical direction $E_1$. Such a conclusion can also be validated by histogram plots counting numbers of models having the same in-plane modulus component $E_1$, $E_2$, or $G_{12}$ for both single material and bi-material woven composite models. Details of the histogram plots and descriptions are shown in Appendix~\ref{App:Woven_Hist}. This many-to-one mapping poses challenges while predicting weave patterns for a given target in-plane modulus (BDP problems), which is later addressed within the Deep Neural Network frameworks.

\subsection{Mechanical Properties of Plain Weave Composites to Other Patterns}
\begin{figure}[h!]
\centering
\subfigure[]{
  \includegraphics[width=0.35\textwidth]{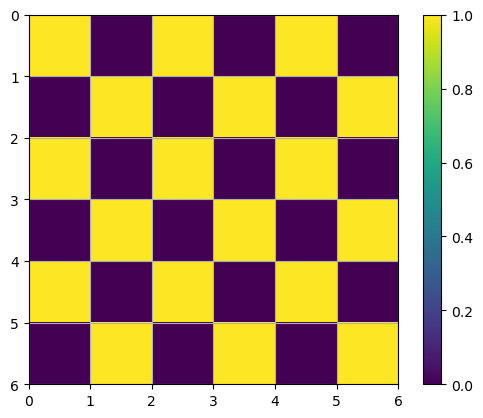}
}
\centering
\subfigure[]{
  \includegraphics[width=0.4\textwidth]{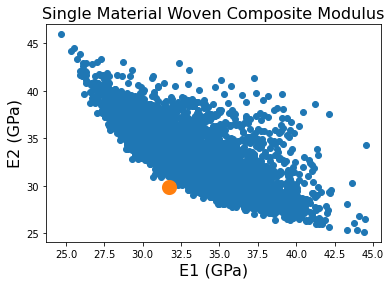}
}
\caption{(a) 6-by-6 representation of the plain weave pattern (b) mechanical properties of plain weave VS all 9000 patterns for single material woven composite (the yellow dot denotes the mechanical properties of plain weave and blue dot denotes the mechanical properties of other weave patterns considered)}
\label{img:plain_weave}
\end{figure}

Among different patterns typically used in woven composites, plain weave, alternating '0' and '1' in its pattern, is the most fundamental weave design in different areas, including aerospace, fashion, and furnishing. However, this does not imply that a plain weave will result in the best mechanical properties. As shown in Figure~\ref{img:plain_weave}, we can see that there are various patterns (28.7\% of the 9000 samples) having better modulus in both $E_1$ and $E_2$ directions compared to plain weave (orange dot). The behavior can be attributed to the plain weave's lowest value of ${n_g}$ for the plain weave, which is discussed in Section \ref{app:WeavePar}. The value of ${n_g}=2$ leads to a maximum number of undulations in both the horizontal and vertical directions, resulting in inferior mechanical properties in both directions. Therefore, it is crucial to explore weave patterns that will result in superior mechanical properties than plain weave.

\subsection{Loss Functions Considered}

This paper considers two types of commonly used loss functions: Mean Squared Error (MSE) and Binary Cross-Entropy (BCE). MSE measures how close the predicted value is to the true value. This paper uses MSE for in-plane modulus-related predictions, defined as Equation~\ref{eqn:mse}.

\begin{equation}
    MSE(y,\Tilde{y}) = \frac{1}{3n} \sum_{i=1}^n\sum_{j=1}^3 (y_{ij} - \Tilde{y}_{ij})^2
\label{eqn:mse}
\end{equation}

Where n is the total sample size, `3' means the size of the in-plane modulus vector, $y_{ij}$ is the predicted value of $i^{th}$ data sample, and $j^{th}$ in-plane modulus. $\Tilde{y}_{ij}$ is the corresponding true value. MSE can be widely used for different prediction tasks. However, it could be a wrong choice for binary classification problems as MSE generally assumes data with normal distribution, while binary classification can be viewed as a Bernoulli distribution. Moreover, the MSE function is non-convex for binary classification problems using activation functions like the Sigmoid function. Thus, we will use BCE defined as Equation~\ref{eqn:bce} for predicting binary woven pattern matrix or binary material sequence vector.

\begin{equation}
\begin{aligned}
    BCE(y,\Tilde{y}) = -\frac{1}{nm} \sum_{i=1}^n\sum_{j=1}^m \Tilde{y}_{ij} log(y_{ij}) + (1-\Tilde{y}_{ij}) log(1-y_{ij})
\end{aligned}
\label{eqn:bce}
\end{equation}

Similar to the definition of MSE, n is the total sample size, and m is the target size. For example, $m=36$ when predicting the 6-by-6 weave pattern and $m=12$ when predicting the 6-by-2 weave material sequence. $y_{ij}$ is the predicted value at $j^{th}$ component in $i^{th}$ model and $\Tilde{y_i}$ is the corresponding true value.

\section{Deep Neural Network Frameworks}

This section will show the detailed Deep Neural Network frameworks we propose to solve the FDP and BDP problems. As briefly mentioned in Section~\ref{intro}, we utilize DCNN to solve the FDP problem, and we propose our PCNN to solve the BDPa and BDPb problems.

\subsection{Forward Direction Prediction: Deep Convolutional Neural Network} \label{sec:FDP_NN}

\begin{figure}[h!]
\centering
	\includegraphics[width=0.98\textwidth]{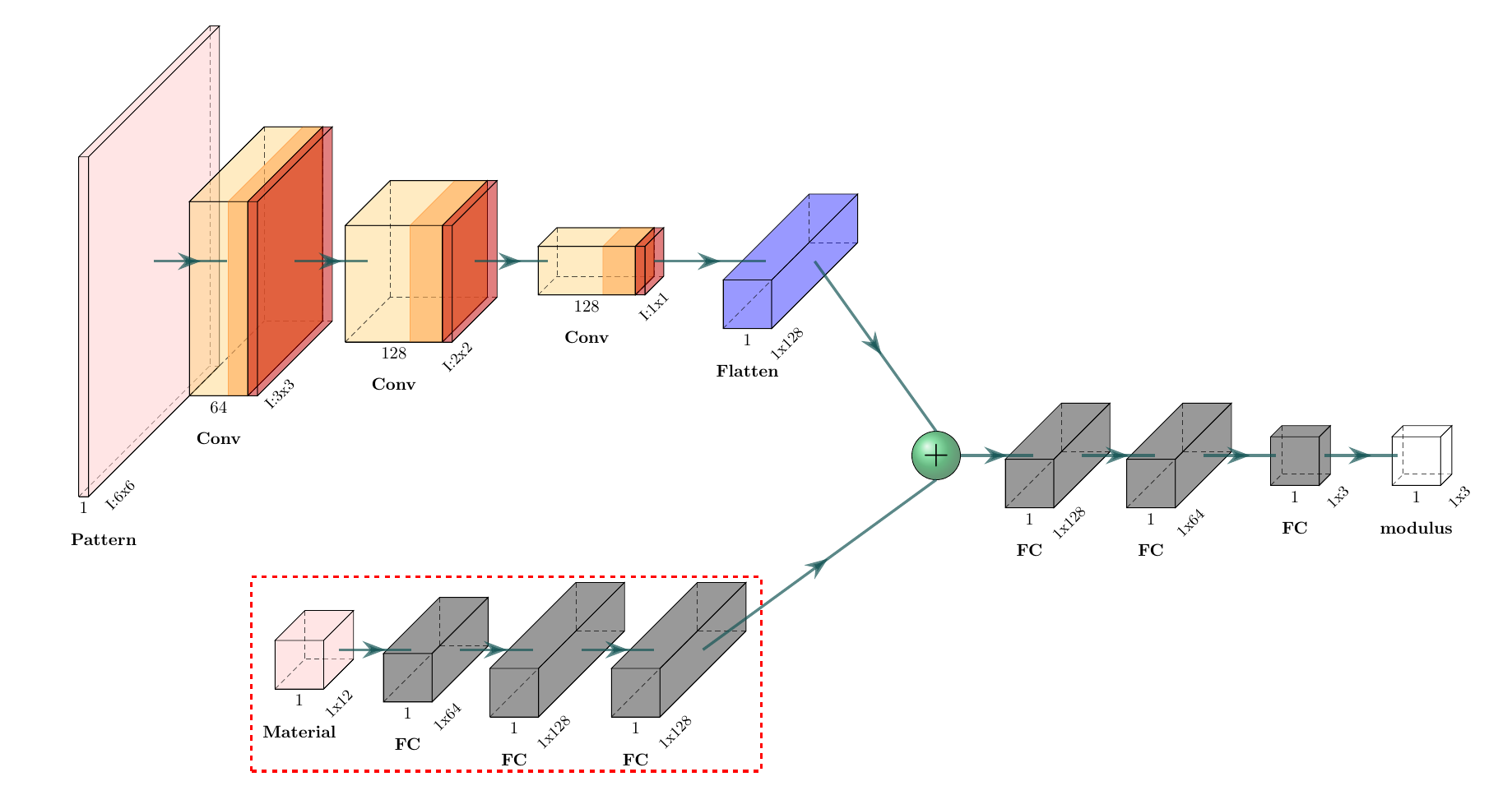}
\caption{Deep Convolutional Neural Network (DCNN) for FDP: pink blocks are the inputs to the Neural Network; orange blocks are convolutional layers with ReLU activation function, and brown blocks are batch-normalization layers following Convolutional layers; the blue block is the Flatten layer that reshapes the input into a vector; gray blocks are Fully Connected layers with ReLU activation function; white blocks are the outputs of the framework. The green ball represents the Concatenation layer. The modules inside the red dashed block are only activated when the material sequence serves as the input for the bi-material woven composite.}
\label{img:Forward_NN}
\end{figure}

For the FDP problem, we developed a Deep Convolutional Neural Network (DCNN), with the overall framework shown in Figure~\ref{img:Forward_NN}. Initially, weave patterns and material sequences are fed into DCNN as inputs. Then, we will use Convolutional layers with ReLU as the activation function for the weave pattern to extract high-level features from the pattern. At the same time, the material assignment vector will be expanded by fully connected layers. Then, extracted features from the weave pattern and material sequence are concatenated into a new feature vector and further used to predict the in-plane modulus through fully connected layers with the ReLU activation function.

\subsection{Backward Direction Prediction: Physics-Constrained Network Framework }

As mentioned in Section~\ref{intro}, the BDP problem is decoupled into two problems: BDPa and BDPb. We have shown there exists a many-to-one mapping, which makes BDP problems much more challenging to handle than FDP problems. This paper proposes two PCNNs for BDPa and BDPb, respectively. Although the two frameworks are slightly different due to different input data, both frameworks are developed based on \textbf{Physics-constraint} using the trained DCNN from the FDP problem to constrain the prediction. 

\subsubsection{Predicting weave pattern from in-plane modulus and material sequence (BDPa)}

\begin{figure}[h!]
\centering
	\includegraphics[width=0.98\textwidth]{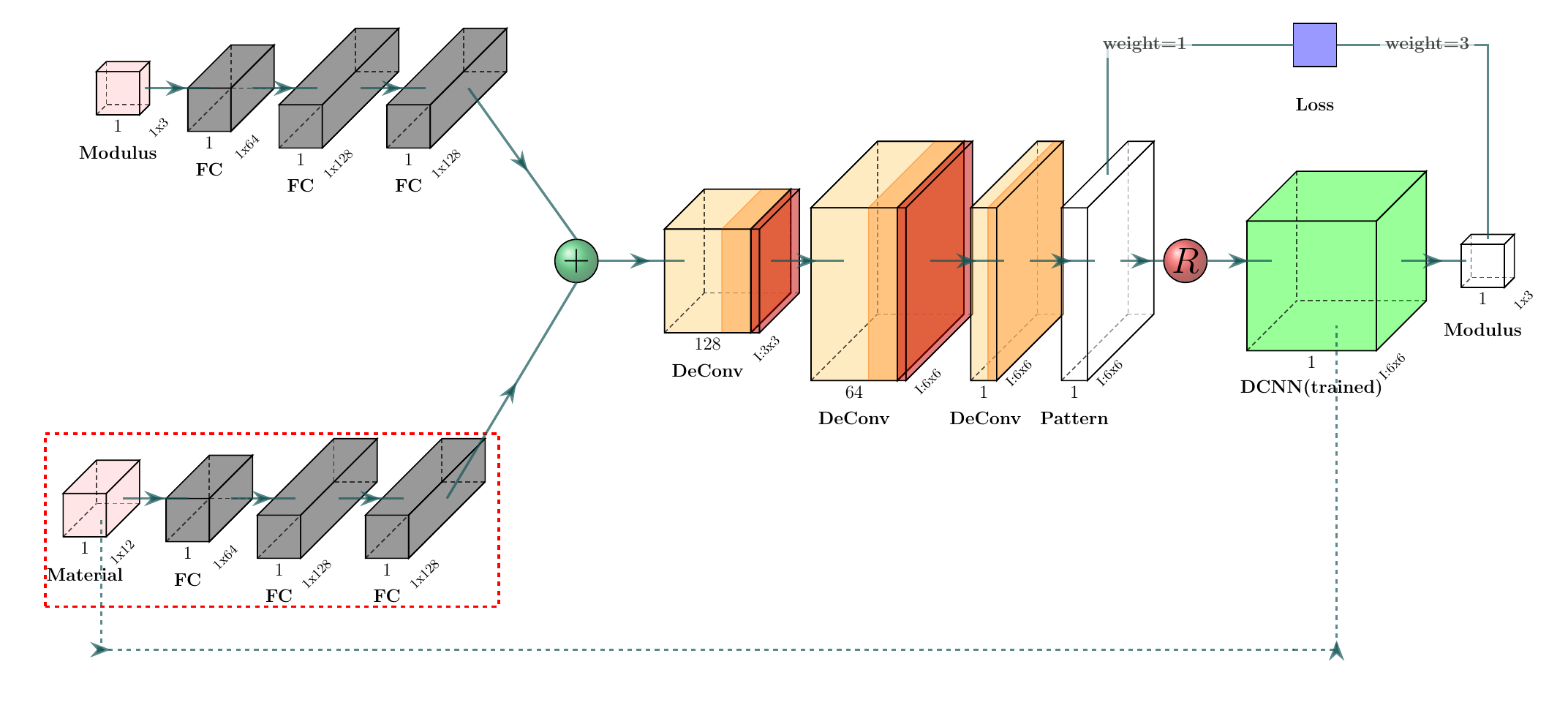}
\caption{Physics-Constraint Neural Network framework for BDPa problem: pink blocks are inputs to the framework; gray blocks are fully connected layers; orange blocks are Deconvolutional layers with LeakyReLU activation function, and brown blocks are batch-normalization layers following the convolutional layers; white blocks are outputs of the framework; the red ball is the rounding layer that rounds the predicted probability vector into binary values to get the binary material vector; the dark green ball is the concatenation layer that concatenates extracted features from in-plane modulus and material assignment; Light green block refers to the previously trained DCNN framework. The modules inside the red dashed block and the green dashed arrow are only activated when material assignment serves as the bi-material woven composite input.}
\label{img:BDPa}
\end{figure}

For single-material woven composite, the BDPa problem is to predict the weave pattern directly from the given in-plane modulus. In contrast, the problem is extended for bi-material woven composite to predict woven patterns from given in-plane modulus and material sequence. The whole framework to solve the BDPa problem is shown in Figure~\ref{img:BDPa}. In-plane modulus and material sequence in pink blocks are the inputs to the framework. The two inputs are expanded through several fully connected layers, concatenated into one vector, and brought into the Deconvolutional layers with LeakyReLU. The deconvolutional layers will expand the feature vector into its original physical space of 6-by-6. Since each weave pattern is a binary matrix, the last Deconvolutional layer uses the Sigmoid activation function. To embed our existing knowledge into the prediction and enhance the prediction accuracy, we add the trained DCNN from Section~\ref{sec:FDP_NN} after the predicted woven pattern and further evaluate the prediction's accuracy in terms of in-plane modulus, as shown in the light green block. To improve the prediction accuracy, we control the weights of modulus-related loss three times larger than the weights of pattern-related loss. For the loss function, the pattern-related loss is calculated based on BCE, and the corresponding modulus-related loss is calculated based on MSE.

\subsubsection{Predicting weave material sequence from in-plane modulus and pattern (BDPb)}

\begin{figure}[h!]
\centering
	\includegraphics[width=0.98\textwidth]{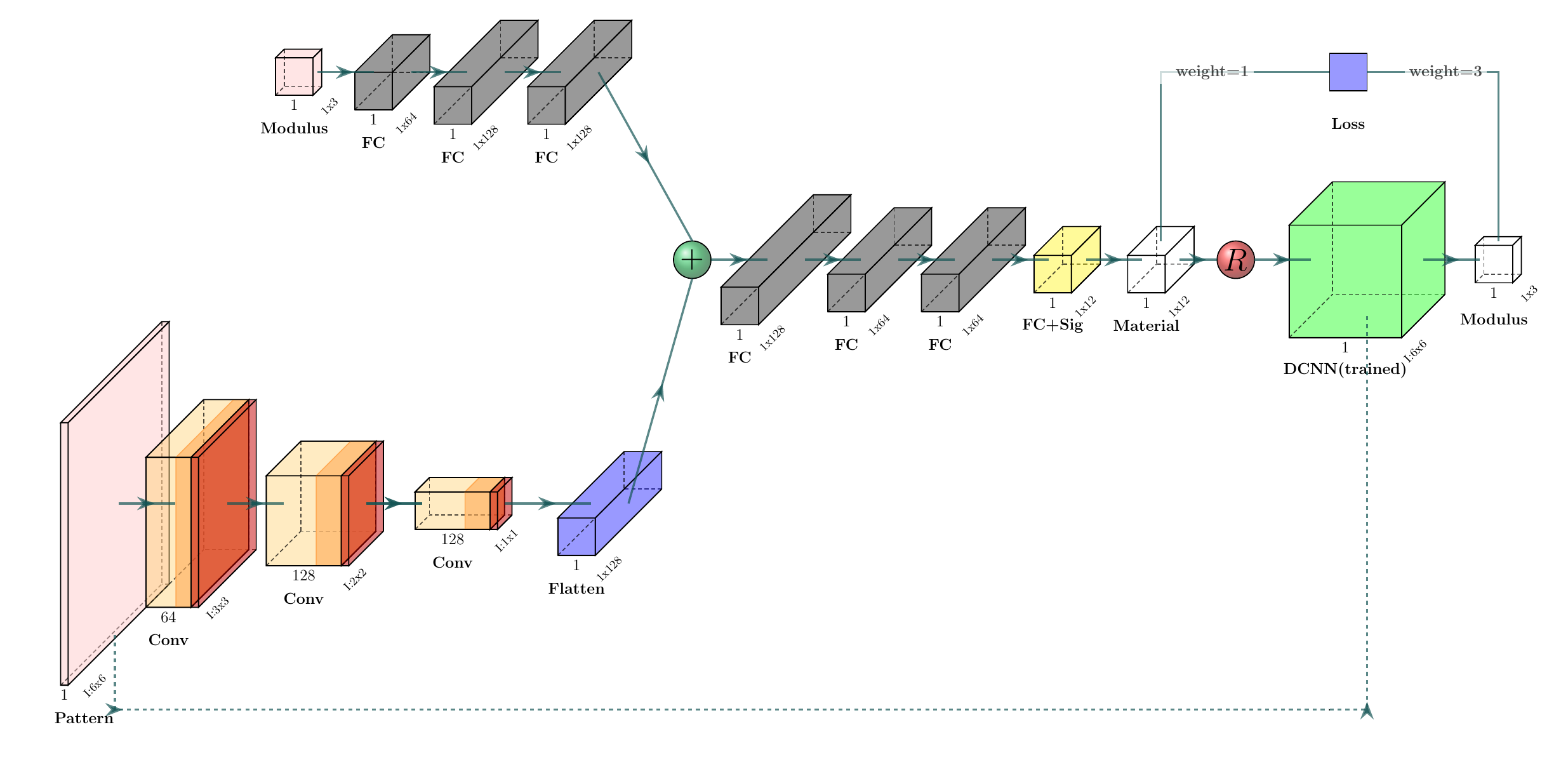}
\caption{Physics-Constraint Neural Network framework for BDPb problem: pick blocks are inputs to the framework; gray blocks are fully connected layers with ReLU as activation function; the yellow block is the fully connected layer with Sigmoid activation function; orange blocks are convolutional layers with ReLU activation function, and brown blocks are batch-normalization layers following the convolutional layers; white blocks are outputs of the framework; the red ball is the rounding layer that rounds the predicted probability vector into binary values to get the binary material vector; the dark green ball is the concatenation layer that concatenates high-level features from in-plane modulus and pattern; light green block refers to the previously trained DCNN framework.}
\label{img:BDPb}
\end{figure}

Compared to the BDPa problem, the BDPb problem focuses on predicting weave material sequence from in-plane modulus and pattern. This framework concentrates only on bi-material woven composites as single-material woven composites have a constant material vector, as shown in Figure~\ref{img:BDPb}. Weave in-plane modulus and pattern serve as inputs to the framework, where the in-plane modulus is passed into several fully connected layers with the ReLU activation function. In contrast, the pattern is passed into several convolutional layers with ReLU activation function followed by batch normalization. The extracted high-level features from the in-plane modulus and pattern are concatenated into a vector and passed into several fully connected layers with the ReLU activation function. Since the material sequence is a binary vector, the last fully connected layer has the sigmoid activation function. Similar to BDPa, to enhance prediction accuracy, we constrain the prediction by adding trained DCNN from Section~\ref{sec:FDP_NN} after the prediction layer. Similar to the BDPa problem, the weights of modulus-related loss are also three times larger than the weights of the material sequence-related loss. For the loss function, the material sequence-related loss is calculated based on BCE, and the corresponding modulus-related loss is calculated based on MSE.

\section{Results and Discussion: Task 1 - Relating Woven Composite Architecture and In-Plane Moduli}
\subsection{Overview of Baseline Models Considered} \label{sec:baseline}
In this research, we use 9000 single-material and 9000 bi-material woven composite models, respectively, to analyze the performance of the proposed Machine Learning frameworks. The data are randomly split into a 60\% training set, 20\% cross-validation set, and 20\% testing set. To control the random split method for comparison, we control the random split seed such that different Machine Learning algorithms are evaluated based on the same data set. 

To evaluate the Machine Learning framework's performance for the FDP problem, we directly assess the in-plane modulus prediction in terms of mean absolute percentage error (MAPE) defined in Equation~\ref{eqn:mape}. 
\begin{equation}
    MAPE=\frac{1}{n}\sum_{t=1}^n |\frac{A_t-F_t}{A_t}|
\label{eqn:mape}
\end{equation}

$A_t$ is the actual value, $F_t$ is the predicted value, and $n$ is the total sample size. On the other hand, since BDP problems are more complex than FDP problems, we will evaluate the prediction error based on MAPE and compare our PCNN performance with other popular baseline models. There are three baseline models considered in this paper: (1) Woven-Decoder, which utilizes the Autoencoder structure\cite{badrinarayanan2017segnet}. Autoencoder framework has been widely used for image-based prediction, like predicting the stress contours\cite{feng2021difference,sepasdar2021data,pang2021dislocation}. The detailed framework of the Woven-Decoder is shown in Appendix~\ref{ED_baseline}. (2) Woven-GAN, which is developed based on the GAN framework. Here we represent the generator using the Woven-Decoder structure while adding the discriminator after the output to classify the output into a binary value. Such binary values will tell if the generator's result is realistic. The GAN-based framework has been used to predict the checkerboard pattern of bi-material composite or to predict the stress distribution contours of different shapes of cantilever beams under certain loading conditions\cite{chen2020generative,jiang2021stressgan}. The GAN and Woven-GAN setup details are shown in Appendix~\ref{GAN_baseline}. (3) Woven-GA, developed based on Genetic Algorithm\cite{whitley1994genetic}. The genetic algorithm is a search heuristic from the theory of natural evolution. It generates new generations through crossover and mutations based on a user-defined fitness function by starting from randomly chosen first generations. A genetic algorithm has been used to determine the complex geometry from targeted mechanical properties, like finding the bi-material composite model design with the highest strength\cite{abueidda2019prediction}. The Woven-GA structure and parameter setup details are shown in Appendix~\ref{GA_baseline}. Although BDPa and BDPb problems have different prediction targets, both problems target finding the best pair of patterns and material sequences to match the target modulus. So both BDPa and BDPb are evaluated based on the MAPE between the target in-plane modulus and the predicted architecture's in-plane modulus.

\subsection{Forward Direction Prediction Results}
As mentioned in Section~\ref{sec:baseline}, the performance of the single-material woven composite is evaluated based on the MAPE values. Since FDP problem aims to predict the in-plane modulus, the MAPE is calculated based on $E_1$, $E_2$, and $G_{12}$, respectively. We will validate the Neural Network's performance on single-material and bi-material woven composite separately. 

Table~\ref{tab:fdp_error} shows the prediction results of the single-material and bi-material woven composites. We see that for single material woven composites, our proposed DCNN's prediction error for $E_1$ and $E_2$ are below 2\%. The prediction error for $G_{12}$ is low as shear modulus does not vary much for single material woven composite, as shown in Figure~\ref{img:single-mat_modulus}. For bi-material woven composites, as each in-plane modulus is more distributed for different models, the prediction error of our proposed DCNN will increase marginally. Our proposed DCNN could achieve prediction error at around 4\% for $E_1$ and $E_2$ and below 2\% for $G_{12}$. Since woven composite in-plane modulus ranges from around $15 GPa \sim 45 GPa$, the average error is around $0.1 \sim 0.2 GPa$. These results indicate that our proposed DCNN effectively represents the relationship between woven architecture and its in-plane modulus.

\begin{table}[h!]
\centering
\caption{FDP prediction error rate}
\resizebox{0.5\textwidth}{!}{
\begin{tabular}{cccc}
\hline
Error Rate                       & $E_1$     & $E_2$     & $G_{12}$    \\ \hline
Single-Material Woven & 1.86\% & 1.89\% & 0.25\% \\ 
Bi-Material Woven     & 4.08\% & 3.76\% & 1.84\% \\ \hline
\end{tabular}
}
\label{tab:fdp_error}
\end{table}

\subsection{Backward Direction Prediction Results}

As discussed, the BDP problem is split into two sub-problems: BDPa and BDPb. To evaluate the performance of our proposed PCNN, we consider the prediction error of BDPa and BDPb problems for single material and bi-material woven composite separately.

\subsubsection{Single material woven composite prediction results}
For single-material woven composite, we compare our proposed Machine Learning framework with three baseline models described in Section~\ref{sec:baseline}. To compare the prediction between different Machine Learning frameworks, we compare the prediction accuracy and duration, as shown in Table~\ref{tab:BDP_single}. From the results, we show that:
\begin{enumerate}
    \item Woven-GA gives the highest prediction accuracy for all models. However, since it is a heuristic searching algorithm, it will take more than one hour for each prediction, and such searching needs to be repeated every time we use it. Also, the performance of heuristic searching largely depends on the data sample. Thus Woven-GA is a costly method and will not be considered.
    \item For the rest of the Deep Neural Network-based models, as the models are learned through training-predicting, it takes much less time for each prediction. Compared to Woven-Decoder and Woven-GAN, our PCNN has significantly reduced the prediction error to around 2\% for $E_1$ and $E_2$ giving the best overall prediction compared to all baseline models.
\end{enumerate}

\begin{table}[h!]
\centering
\caption{BDPa prediction error rate for single material woven composite}
\resizebox{0.6\textwidth}{!}{
\begin{tabular}{ccccc}
\hline
Error Rate    & $E_1$  & $E_2$     & $G_{12}$    & Prediction Time \\ \hline
Woven-Decoder & 7.87\% & 7.26\% & 0.33\% & \textless\textless 1sec  \\ 
Woven-GAN     & 4.34\% & 5.27\% & 0.31\% & \textless\textless 1sec  \\ 
Woven-GA      & 0.05\% & 0.01\% & 0.58\% & 60 mins 18 secs \\ 
PCNN    & 2.38\% & 1.72\%   & 0.31\% & \textless\textless 1sec  \\ \hline
\end{tabular}
}
\label{tab:BDP_single}
\end{table}

\subsubsection{Bi-material woven composite prediction results}
Since BDP problems for bi-material woven composite consist of three inputs, prediction with Woven-GA will be even more expensive and will not be considered. Table~\ref{tab:BDPa_bi} shows the prediction results for the BDPa problem, comparing our proposed model and the other two baseline models. Figure~\ref{img:BDPa_pattern_predict} shows images of the predicted woven patterns for a given in-plane modulus and material sequence. Based on the analysis results, we notice that:
\begin{enumerate}
    \item Compared to baseline models, PCNN significantly reduces the error rate of $E_1$ and $E_2$ predictions from around 10\% to 3.6\%, and the error rate of $G_{12}$ also decreases to around 1.3\%. Thus, PCNN outperforms the baseline models considered.
    \item For the predicted pattern, we can find that PCNN gives the closest prediction to the original weave pattern. Furthermore, we show a detailed quantitative explanation of why PCNN is superior to other models in Section~\ref{sec:NN_pred_GLCM} using our proposed GLCM-based feature (in Section~\ref{sec:glcm_opt}) analysis.
\end{enumerate}

\begin{table}[h!]
\centering
\caption{BDPa prediction error rate for bi-material woven composite}
\resizebox{0.65\textwidth}{!}{
\begin{tabular}{ccccc}
\hline
Error Rate    & $E_1$      & $E_2$      & $G_{12}$     & Prediction Time \\ \hline
Woven-Decoder & 9.31\%  & 9.45\%  & 5.01\%  & \textless \textless 1sec  \\ 
Woven-GAN     & 10.83\% & 11.71\% & 10.62\% & \textless \textless 1sec  \\ 
PCNN    & 3.60\%  & 3.71\%  & 1.34\%  & \textless \textless 1sec  \\ \hline
\end{tabular}
}
\label{tab:BDPa_bi}
\end{table}

\begin{figure}[h!]
\centering
\subfigure[]{
  \includegraphics[width=0.23\textwidth]{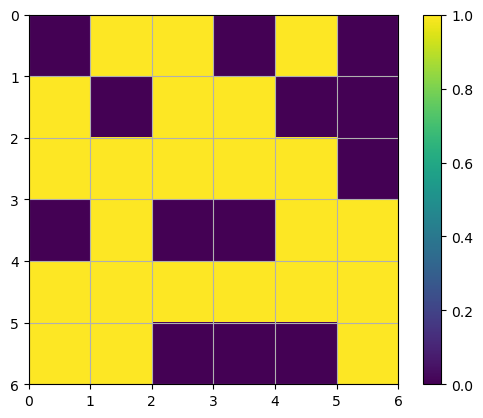}
}
\subfigure[]{
  \includegraphics[width=0.23\textwidth]{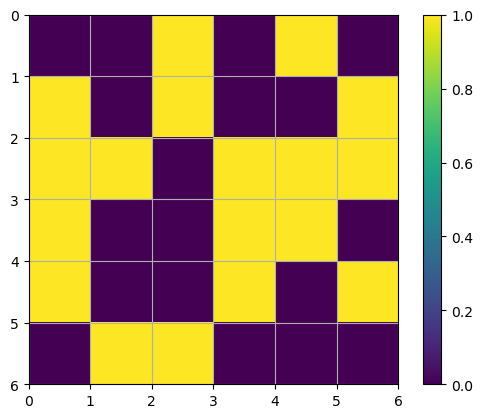}
}
\centering
\subfigure[]{
  \includegraphics[width=0.23\textwidth]{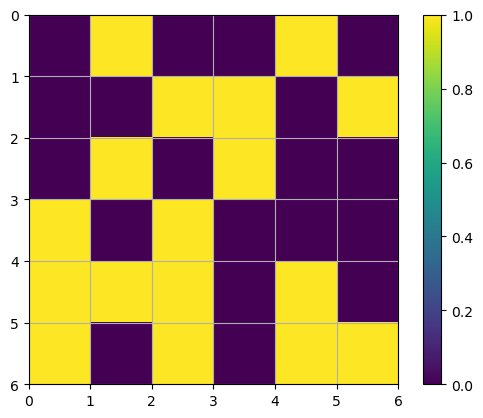}
}
\centering
\subfigure[]{
  \includegraphics[width=0.23\textwidth]{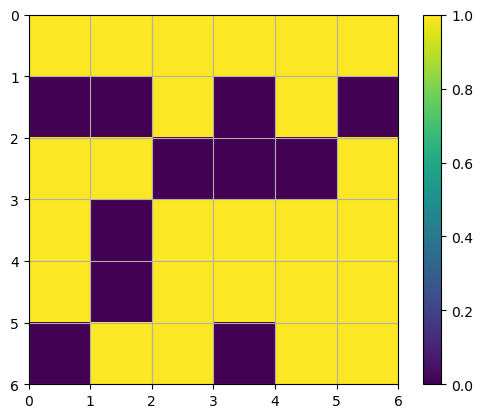}
}
\caption{Predicted bi-material weave pattern for BDPa problem: (a) original weave pattern (b) predicted weave pattern from Woven-Decoder (c) predicted weave pattern from Woven-GAN (d) predicted weave pattern from PCNN}
\label{img:BDPa_pattern_predict}
\end{figure}

We next evaluate the performance of different models for BDPb problems. Table~\ref{tab:BDPb_bi} shows the prediction result for the BDPb problem, comparing our proposed and baseline models. From these results, we observe that compared to baseline models, Woven-Decoder and Woven-GAN, our proposed PCNN could vastly reduce the prediction error from above 10\% to around 5\% for all three in-plane moduli. Consequently, we can conclude that for both BDPa and BDPb problems, our proposed PCNN can significantly improve the prediction accuracy for all three in-plane moduli.

\begin{table}[h!]
\centering
\caption{BDPb prediction error rate for bi-material woven composite}
\resizebox{0.65\textwidth}{!}{
\begin{tabular}{ccccc}
\hline
Error Rate    & $E_1$      & $E_2$      & $G_{12}$     & Prediction Time \\ \hline
Woven-Decoder & 11.74\%  & 11.73\%  & 11.50\%  & \textless \textless 1sec  \\
Woven-GAN     & 15.28\%  & 12.35\%  & 26.99\%  & \textless \textless 1sec  \\ 
PCNN    &  5.53\%   & 5.65\%   & 4.10\%  & \textless \textless 1sec  \\ \hline
\end{tabular}
}
\label{tab:BDPb_bi}
\end{table}

\subsubsection{Weave pattern modification by bound relaxation of modulus (for manufacturing purpose)}

When predicting the weave pattern in the BDPa problem, we do not add constraints to the predicted pattern. However, during manufacturing, it is usually challenging to weave patterns with continuous yarns or fiber bundles running along the warp and weft directions without an area of interlacing. Solving this issue requires using pre-preg tapes made from "pre-impregnated" fibers and a partially cured polymer matrix. Alternatively, stitching of fibers is needed to maintain the structural integrity of the fabric during manufacturing. Since this process can be time-consuming and expensive, finding weave patterns that do not have continuous yarns (that is, with interlaced region) is essential. To solve this problem, we propose to find weave patterns by modifying the target modulus within specific ranges, which we call \textbf{Modulus Bound Relaxation}.

The expression of Modulus Bound Relaxation can be represented as Equation~\ref{eqn:mod_relax}.

\begin{equation}
    M_{new} = M_{old} + R \times B 
\label{eqn:mod_relax}
\end{equation}

Where $M_{old}$ is the target modulus vector containing $E_1, E_2, G_{12}$, and $M_{new}$ is the updated new modulus vector. $R\in[-1,1]^{d=3}$ is a 3-by-1 vector, with each component randomly generated between -1 and 1. $B$ is a range of scaling factor of $R$, the upper and lower bound of $B$ can be specified by the user. $R \times B$ determines the maximum relaxation we want for the target modulus vector. We linearly increase the value of $B$ from its lower bound to its upper bound to increase the relaxation until we find a weave pattern without continuous yarn. With this method, we can find a surrogate weave pattern with a modulus vector slightly different than our target, but with no continuous yarn issue -- an example of how the Modulus Bound Relaxation algorithm works is shown in Appendix~\ref{sec:mod_relax}.


\subsection{PCNN Performance with Small Dataset}
This section tests how PCNN performs when the dataset is small. Specifically, we analyze the performance of PCNN when we select 3000 and 6000 bi-material woven composite samples each from the 9000 dataset. Table~\ref{tab:BDPa_3000} and Table~\ref{tab:BDPa_6000} show the prediction error for 3000 and 6000 samples. We observe consistent results compared to the previous analysis using 9000 samples from the results. Thus we conclude that for both BDPa and BDPb problems, PCNN provides better and more stable prediction accuracy than the two baseline models.

\begin{table}[h!]
\centering
\caption{Prediction error rate with 3000 bi-material woven composite samples}
\resizebox{0.7\textwidth}{!}{
\begin{tabular}{cccccc}
\hline
                                                                        &   Error Rate            & $E_1$      & $E_2$      & $G_{12}$     & Prediction Time           \\ \hline
\multirow{3}{*}{\begin{tabular}[c]{@{}c@{}}BDPa\\ Problem\end{tabular}} & Woven-Decoder & 10.61\% & 11.97\% & 10.30\% & \textless{}\textless 1sec \\ 
                                                                        & Woven-GAN     & 8.67\%  & 6.27\%  & 6.87\%  & \textless{}\textless 1sec \\ 
                                                                        & PCNN    & 4.23\%  & 4.28\%  & 3.63\%  & \textless{}\textless 1sec \\ \hline
\multirow{3}{*}{\begin{tabular}[c]{@{}c@{}}BDPb\\ Problem\end{tabular}} & Woven-Decoder & 8.73\%  & 8.84\%  & 10.93\% & \textless{}\textless 1sec \\  
                                                                        & Woven-GAN     & 11.53\% & 12.55\% & 11.65\% & \textless{}\textless 1sec \\ 
                                                                        & PCNN    & 4.69\%  & 3.70\%  & 1.53\%  & \textless{}\textless 1sec \\ \hline
\end{tabular}
}
\label{tab:BDPa_3000}
\end{table}

Furthermore, Table~\ref{tab:BDPa_6000} shows the prediction error of BDPa and BDPb problems under the 6000 sample. From the result, we can validate that for both BDPa and BDPb problems, PCNN has much better prediction accuracy compared to the two baseline models.

\begin{table}[h!]
\centering
\caption{Prediction error rate with 6000 bi-material woven composite samples}
\resizebox{0.7\textwidth}{!}{
\begin{tabular}{cccccc}
\hline
                                                                        & Error Rate              & $E_1$      & $E_2$      & $G_{12}$     & Prediction Time           \\ \hline
\multirow{3}{*}{\begin{tabular}[c]{@{}c@{}}BDPa\\ Problem\end{tabular}} & Woven-Decoder & 11.61\% & 12.20\% & 12.33\% & \textless{}\textless 1sec \\  
                                                                        & Woven-GAN     & 6.48\%  & 6.91\%  & 6.14\%  & \textless{}\textless 1sec \\  
                                                                        & PCNN    & 3.81\%  & 3.93\%  & 3.18\%  & \textless{}\textless 1sec \\ \hline
\multirow{3}{*}{\begin{tabular}[c]{@{}c@{}}BDPb\\ Problem\end{tabular}} & Woven-Decoder & 9.24\%  & 9.43\%  & 10.99\% & \textless{}\textless 1sec \\  
                                                                        & Woven-GAN     & 12.20\% & 11.56\% & 10.24\% & \textless{}\textless 1sec \\  
                                                                        & PCNN    & 4.10\%  & 4.20\%  & 1.49\%  & \textless{}\textless 1sec \\ \hline
\end{tabular}
}
\label{tab:BDPa_6000}
\end{table}

\section{Results and Discussion: Task 2 - Feature-based Statistical Representation} \label{sec:glcm_opt}

In the previous sections, we proposed DCNN and PCNN to establish the bridge between woven architectures and the corresponding modulus. Our proposed Deep Learning frameworks deliver better predictions for FDP, BDPa, and BDPb problems than baseline models. However, the high-level features extracted by PCNN are challenging to understand and be used for other tasks like optimization. Thus, we want to know what physically or statistically meaningful features control the woven composite in-plane modulus, and how we can use these features to optimize the woven composite. To that end, we conduct the GLCM-based feature analysis.

\subsection{Statistical Features from Weave Pattern} \label{sec:pat_stat_feature}
Since weave patterns are represented by a checkerboard model, we considered this as a type of texture with pixel values of '0' or '1'. Texture features describe the spatial distribution of pixels (cells) that reflect an object's roughness, smoothness, granularity, and randomness. Texture can also be used to segment images into regions of interest and classify those regions into regular texture and quasi-regular texture. Regular texture's element follows a specific pattern, whereas quasi-regular texture's element has an arbitrary shape and is distributed based on intensity. Standard texture feature extraction methods include statistical, structural, and spectral methods. This paper utilizes the statistical method and constructs the Gray Level Co-occurrence Matrices (GLCM). GLCM elements are defined in Equation~\ref{eqn:GLCM}.

\begin{equation}
    C_{\Delta x, \Delta y}(i,j) = \sum_{x=1}^n \sum_{y=1}^m \mathbbm{1}_{[I(x,y)=i, I(x+\Delta x, y+\Delta y)=j]}
\label{eqn:GLCM}
\end{equation}

where, I is the grey-level image, $i$ and $j$ are pixel values. $n,m$ is the size of image, ($x,y$) is the starting position, and ($\Delta x$, $\Delta y$) represent the offset from starting position. As our checkerboard models are binary matrices in this paper, the GLCM will be a 2-by-2 matrix, where we consider transitions of $0\rightarrow0$, $0\rightarrow1$, $1\rightarrow0$, and $1\rightarrow1$. Further, we consider four different directions (horizontal, vertical, and two diagonal directions) during GLCM calculation. The texture features considered are contrast, correlation, energy, and homogeneity. Definition of each statistical term is summarized in Appendix~\ref{app:tex-eqn}, where we refer to Beyer~\cite{hall2017glcm}, Haralick~\cite{haralick1992computer}, and Bevk~\cite{bevk2002statistical}.
Thus, for each woven composite pattern matrix, we will extract $4\times4=16$ features. We can achieve several excellent properties by extracting statistical features from 2-by-2 GLCM in the form $\begin{bmatrix}
    a & b \\
    c & d
\end{bmatrix}$,  leading to the following propositions:

\begin{Proposition}
Weave pattern GLCM statistical features (Contrast, Correlation, Energy, Homogeneity) correspond to a unique 2-by-2 GLCM.
\label{claim:1}
\end{Proposition}

\begin{Proposition}
Weave pattern GLCM's Energy (with the help of several other statistical features) tells the relative relationship between different pattern transitions ($0\rightarrow0$, $0\rightarrow1$, $1\rightarrow0$, and $1\rightarrow1$) in the physical space.
\label{claim:2}
\end{Proposition} 

\begin{Proposition}
Weave pattern GLCM's Contrast and Homogeneity tell the frequencies of homogeneous transition ($0\rightarrow0,1\rightarrow1$) and in-homogeneous transition ($0\rightarrow1,1\rightarrow0$).
\end{Proposition}

Proofs of the above propositions can be found in Appendix~\ref{sec:glcm_proof}.

\subsection{Statistical Features from Weave Material Sequence} \label{sec:mat_stat_feature}
Statistical features from woven composite material are only considered for bi-material weaves, as the material sequence for a single material is a uniform constant vector. We split the material vector into the vector for the weft and the vector for the warp. First, we extract statistical features for each material vector, including mean, median, and standard deviation from the vector. For each material vector of woven composite, we extract six features. Specifically, as material vectors can be constant, we do not include skewness and kurtosis in this study. Then, to account for the sequence information, we propose another statistical parameter called Vector Energy (VE): for any vector $V$, the vector energy is defined as $VE = \sum_{i=1}^L i*V(i)$. Here $L$ denotes the length of the vector, and $V(i)$ is the value of the i-th component in $V$.


\subsection{Regression Analysis of Extracted Features} \label{sec:regress_feature}
To understand whether each statistical feature is positively or negatively correlated with in-plane modulus, we use regression analysis to determine the weights of each feature. Specifically, we consider two different cases: (1) how each feature is correlated with the label of the composite model. We group the woven composite models into two groups: one group with a better overall modulus ($E_{all}=E_1+E_2+G_{12}$) is labeled as `1,' and the other group is labeled as '0'; (2) how each feature is correlated with the value of individual in-plane modulus. Here, the better overall modulus is selected based on the highest quantiles across all datasets under consideration.

\subsubsection{Regression analysis on the overall modulus of model}  \label{sec:data_group}

To understand what features contribute to woven composite's better overall modulus, we utilize the Ridge regression to predict the overall modulus directly from the target statistical features. The regression model is defined as Equation~\ref{eqn:ridge}.

\begin{equation}
    \min_{\bold{w},\lambda} ||\bold{y}-X\bold{w}||^2+\lambda ||\bold{w}||^2
    \label{eqn:ridge}
\end{equation}

where $X$ is the matrix formed by feature vectors. $y$ is vector containing $E_{all}$ values. Similarly, values in vector $\bold{w}$ tell us how each statistical feature is correlated to the woven composite model's $E_{all}$ values.

\subsubsection{Weave pattern feature analysis}
We first fix the weave material sequence and analyze how the weave pattern features are correlated with the overall composite modulus $E_{all}$. From the regression analysis results, we observe that the two regression models, one each for single and bi-material, have weights with the same signs, as shown in Table~\ref{tab:sign_glcm}. From the results, we conclude that contrast and correlation are negatively correlated with the woven composite's overall modulus, while energy and homogeneity are positively correlated. This trend can be validated by case studies shown later in Sections ~ \ref{sec:pat_case_2} and \ref{sec:pat_case_3}, where we show that each GLCM feature controls the woven composite's modulus as indicated by the regression results. Furthermore, we prove that the GLCM features can be used to optimize weave patterns and guide the woven composite design by a case studies in Sections ~ \ref{sec:pat_case_2} and \ref{sec:pat_case_3}.

The regression analysis found that the homogeneity is positively correlated, whereas the contrast is negatively correlated. These results also agree with the analytical models discussed in Section~\ref{app:WeavePar}. If we subtract the value of contrast from homogeneity with a specific multiplication constant, we obtain the value of homogeneous transitions in weave patterns ($0\rightarrow0,1\rightarrow1$). Therefore, according to regression analysis, we can conclude that increasing the number of homogeneous transitions will result in higher in-plane moduli. This can be attributed to fewer undulations in the weave patterns with introductions of more homogeneous transitions. 

\begin{table}[h!]
\centering
\caption{Sign of weights for weave pattern features}
\resizebox{0.9\textwidth}{!}{
\begin{tabular}{ccccccccc}
\hline
                & \multicolumn{1}{c}{Contrast} & \multicolumn{1}{c}{Correlation} & \multicolumn{1}{c}{Energy} & Homogeneity & \multicolumn{1}{c}{Contrast} & \multicolumn{1}{c}{Correlation} & \multicolumn{1}{c}{Energy} & Homogeneity \\ \hline
                & \multicolumn{4}{c}{GLCM 1}                                                                                  & \multicolumn{4}{c}{GLCM 2}                                                                                  \\ 
Single Material & \multicolumn{1}{c}{-}        & \multicolumn{1}{c}{-}           & \multicolumn{1}{c}{+}      & +           & \multicolumn{1}{c}{-}        & \multicolumn{1}{c}{-}           & \multicolumn{1}{c}{+}      & +           \\ 
Bi-Material     & \multicolumn{1}{c}{-}        & \multicolumn{1}{c}{-}           & \multicolumn{1}{c}{+}      & +           & \multicolumn{1}{c}{-}        & \multicolumn{1}{c}{-}           & \multicolumn{1}{c}{+}      & +           \\ \hline
                & \multicolumn{4}{c}{GLCM 3}                                                                                  & \multicolumn{4}{c}{GLCM 4}                                                                                  \\
Single Material & \multicolumn{1}{c}{-}        & \multicolumn{1}{c}{-}           & \multicolumn{1}{c}{+}      & +           & \multicolumn{1}{c}{-}        & \multicolumn{1}{c}{-}           & \multicolumn{1}{c}{+}      & +           \\ 
Bi-Material     & \multicolumn{1}{c}{-}        & \multicolumn{1}{c}{-}           & \multicolumn{1}{c}{+}      & +           & \multicolumn{1}{c}{-}        & \multicolumn{1}{c}{-}           & \multicolumn{1}{c}{+}      & +           \\ \hline
\end{tabular}
}
\label{tab:sign_glcm}
\end{table}

\subsubsection{Weave material feature analysis} \label{sec:mat_feature_ana}

We further consider how the weave material features correlate with its overall modulus $E_{all}$ by fixing the weave pattern. Here, we consider two fixed weave patterns as shown in Figure~\ref{img:woven_mat_study_pat}. Each weave pattern is combined with 500 randomly distributed binary material vectors for regression analysis. Similarly, we consider the one proposed regression model that corresponds to the two fixed weave patterns. 

The regression analysis results show that the mean value of material vector significantly controls the overall modulus of the woven composites. This implies that the sequence of the material vector is much less critical than the number of different materials on the property. This conclusion is validated through case studies later in Section~\ref{sec:case_mat_1} and \ref{sec:case_mat_2} with different mean values of the material sequence. We have also shown that for the two patterns considered, we want to increase the ratio between material `1' and material `0' in vertical yarns and decrease it in horizontal yarns. Therefore, we can optimize the overall modulus for a given weave pattern by varying the mean value of the material vector followed by the sequence.

\section{Results and Discussion: Task 3 - Feature-based Statistical Optimization Case Studies}

After proposing the GLCM representation strategy in Section~\ref{sec:glcm_opt}, in this section, we demonstrate how it can be effectively used for optimization by applying it to different weave models. Specifically, we focus on optimizing the weave pattern given a material sequence and optimizing the material sequence given a weave pattern.

\subsection{Weave Pattern Optimization}
In this section, we primarily investigate the effectiveness of our GLCM feature-based optimization approach for improving weave patterns through several case studies. Firstly, we compare our optimization strategy with the baseline model GIDN. Then, we utilize the GLCM to gain insights into why our PCNN produces better predictions and how GLCM features can be effectively employed to optimize the weave pattern.

\subsubsection{Weave pattern optimization compared with baseline model - GIDN}
We first compare our GLCM feature-based optimization with a baseline model GIDN\cite{chen2020generative} mentioned in Section~\ref{intro}. GIDN consists of a designer and predictor and aims to overcome the issue of local minima by using random initialization based on a Gaussian distribution. The authors claimed in their paper that GIDN, with 1000 randomly initialized models, can find the optimal design. To demonstrate that our GLCM method proposes superior weave models compared to GIDN, we consider GIDN with an increased number of random initialization samples, namely GIDN-1000, GIDN-2000, and GIDN-5000. To compare the performance of the GLCM model and GIDN, we randomly select a weave pattern and a material sequence each time and attempt to find the optimal weave model using GLCM features and different variants of GIDN. We present several case studies, and the optimization results are summarized in Table~\ref{tab:gidn_glcm_1}, Table~\ref{tab:gidn_glcm_2}, and Table~\ref{tab:gidn_glcm_3}. The '-N' means 'normalized' as all tables demonstrate that the GLCM-based optimization method outperforms the GIDN models. Furthermore, since the GLCM-based method can directly transform from the GLCM space back to the original physical space, it is significantly more efficient than the search-based GIDN framework.

\begin{table}[h!]
\centering
\caption{Test Case 1: Optimization for weave model with material sequence [1,0,1,0,1,1,0,0,1,0,0,0]}
\begin{tabular}{cccccc}
\hline
    & Original Model  & GIDN-1000  & GIDN-2000  & GIDN-5000 & GLCM  \\ \hline
E1-N  & 33.73           & 29.23      & 25.72      & 30.69     & 39.80 \\
E2-N  & 21.85           & 31.85      & 36.25      & 33.93     & 24.67 \\
G12-N & 25.02           & 25.88      & 25.85      & 25.32     & 27.40 \\
Sum & 80.60           & 86.96      & 87.82      & 89.94     & 91.87 \\ \hline
\end{tabular}
\label{tab:gidn_glcm_1}
\end{table}

\begin{table}[h!]
\centering
\caption{Test Case 2: Optimization for weave model with material sequence [1,1,0,0,1,1,1,1,0,0,1,1]}
\begin{tabular}{cccccc}
\hline
    & Original Model  & GIDN-1000  & GIDN-2000  & GIDN-5000 & GLCM  \\ \hline
E1-N  & 31.85          & 24.84     & 25.41     & 31.83     & 38.23 \\
E2-N  & 18.99          & 30.18     & 30.94     & 25.33     & 22.43 \\
G12-N & 22.55          & 22.31     & 22.88     & 23.32     & 25.52 \\
Sum   & 73.39          & 77.33     & 79.23     & 80.48     & 86.18 \\ \hline
\end{tabular}
\label{tab:gidn_glcm_2}
\end{table}

\begin{table}[h!]
\centering
\caption{Test Case 3: Optimization for weave model with material sequence [1,1,1,1,1,1,0,0,0,0,0,0]}
\begin{tabular}{cccccc}
\hline
    & Original Model  & GIDN-1000  & GIDN-2000  & GIDN-5000 & GLCM  \\ \hline
E1-N  & 31.53           & 30.22      & 30.17      & 30.98     & 37.63 \\
E2-N  & 21.50           & 31.14      & 32.35      & 31.14     & 24.08 \\
G12-N & 23.87           & 24.38      & 24.61      & 24.69     & 26.13 \\
Sum   & 76.90           & 85.74      & 87.13      & 87.61     & 87.84 \\ \hline
\end{tabular}
\label{tab:gidn_glcm_3}
\end{table}

\subsubsection{Comparing pattern prediction from different Neural Network frameworks based on GLCM features}\label{sec:NN_pred_GLCM}

Figure~\ref{img:BDPa_pattern_predict} illustrates the predicted patterns generated by different Neural Network frameworks for a bi-material woven composite with the same material sequence and in-plane modulus. At first glance, the PCNN prediction appears closer to the original pattern. However, considering the many-to-one mapping discussed in Section~\ref{sec:many-to-one}, we aim to perform a more analytical comparison of the results. We employ our proposed GLCM-based feature analysis approach to achieve this by converting these predicted patterns into GLCMs. Subsequently, we will compare the statistical features of these GLCMs, as they can effectively represent the in-plane modulus.

\begin{figure}[h!]
\centering
\subfigure[]{
  \includegraphics[width=0.23\textwidth]{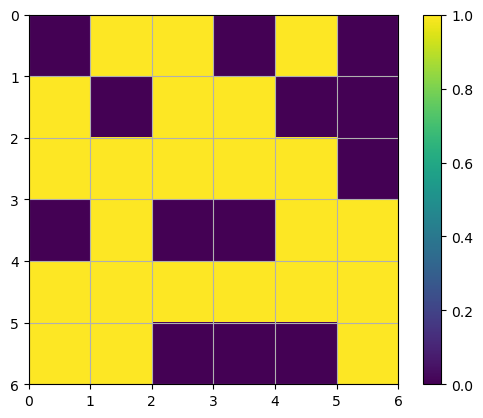}
}
\subfigure[]{
  \includegraphics[width=0.23\textwidth]{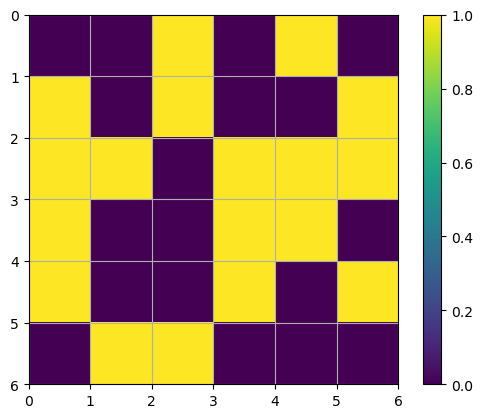}
}
\centering
\subfigure[]{
  \includegraphics[width=0.23\textwidth]{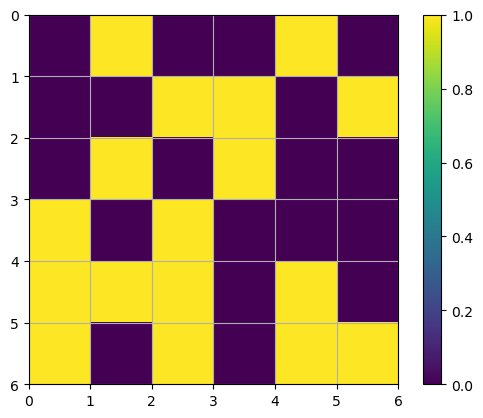}
}
\centering
\subfigure[]{
  \includegraphics[width=0.23\textwidth]{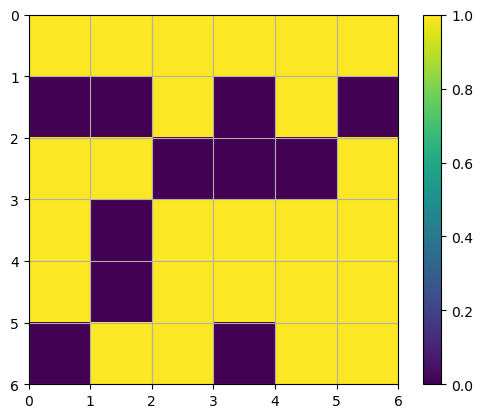}
}
\caption{Predicted bi-material weave pattern for BDPa problem: (a) original weave pattern (b) predicted weave pattern from Woven-Decoder (c) predicted weave pattern from Woven-GAN (d) predicted weave pattern from PCNN}
\label{img:BDPa_pattern_predict}
\end{figure}

We consider GLCM in all four directions: horizontal, vertical, $45^{\circ}$ and $-45^{\circ}$. The corresponding GLCM statistical features are summarized in Table~\ref{tab:glcm_pred_pattern}. From the regression analysis, we know the weights of the statistical features are at the same level, so we can roughly estimate the closeness of different feature vectors by the L2-norm of their differences. By calculating the corresponding L2-norm, we see that: $||F_o-F_d||_2=0.629$, $||F_o-F_g||_2=0.668$ and $||F_o-F_p||_2=0.416$, where $F_o$ denotes the feature vector of the original pattern, $F_d$ from the Woven-Decoder, $F_g$ from the Woven-GAN, and $F_p$ from the PCNN. Furthermore, we can notice that the feature vector of PCNN is closer to the original pattern than in other frameworks.

\begin{table}[h!]
\centering
\caption{GLCM statistical features of predicted weave patterns}
\resizebox{0.99\textwidth}{!}{
\begin{tabular}{ccccccccc}
\hline
                 & \multicolumn{1}{c}{Correlation} & \multicolumn{1}{c}{Contrast} & \multicolumn{1}{c}{Energy}  & Homogeneity & \multicolumn{1}{c}{Correlation} & \multicolumn{1}{c}{Contrast} & \multicolumn{1}{c}{Energy} & Homogeneity \\ \hline
                 & \multicolumn{4}{c}{GLCM 1}                                                                                   & \multicolumn{4}{c}{GLCM 2}                                                                                  \\ 
Original Pattern & \multicolumn{1}{c}{0.4333}      & \multicolumn{1}{c}{-0.0476}  & \multicolumn{1}{c}{0.2994}  & \multicolumn{1}{c}{0.7833}      & \multicolumn{1}{c}{0.5333}      & \multicolumn{1}{c}{-0.2000}  & \multicolumn{1}{c}{0.3067} & 0.7333      \\ \
Woven-Decoder    & \multicolumn{1}{c}{0.6333}      & \multicolumn{1}{c}{-0.2681}  & \multicolumn{1}{c}{-0.2683} & 0.6833      & \multicolumn{1}{c}{0.5333}      & \multicolumn{1}{c}{-0.0714}  & \multicolumn{1}{c}{0.2533} & 0.7333      \\ 
Woven-GAN        & \multicolumn{1}{c}{0.7000}      & \multicolumn{1}{c}{-0.4016}  & \multicolumn{1}{c}{0.2906}  & 0.6500      & \multicolumn{1}{c}{0.5333}      & \multicolumn{1}{c}{-0.0714}  & \multicolumn{1}{c}{0.2533} & 0.7333      \\ 
PCNN       & \multicolumn{1}{c}{0.4333}      & \multicolumn{1}{c}{0.1086}   & \multicolumn{1}{c}{0.2683}  & 0.7833      & \multicolumn{1}{c}{0.6667}      & \multicolumn{1}{c}{-0.3393}  & \multicolumn{1}{c}{0.2800} & 0.6667      \\ \hline
                 & \multicolumn{4}{c}{GLCM 3}                                                                                   & \multicolumn{4}{c}{GLCM 4}                                                                                  \\ 
Original Pattern  & \multicolumn{1}{c}{0.4000}      & \multicolumn{1}{c}{0.0809}   & \multicolumn{1}{c}{0.3248}  & 0.8000      & \multicolumn{1}{c}{0.5600}      & \multicolumn{1}{c}{-0.2868}  & \multicolumn{1}{c}{0.3184} & 0.7200      \\ 
Woven-Decoder      & \multicolumn{1}{c}{0.5200}      & \multicolumn{1}{c}{-0.0400}  & \multicolumn{1}{c}{0.2504}  & 0.7400      & \multicolumn{1}{c}{0.4400}      & \multicolumn{1}{c}{0.1200}   & \multicolumn{1}{c}{0.2536} & 0.7800      \\ 
Woven-GAN        & \multicolumn{1}{c}{0.4000}      & \multicolumn{1}{c}{0.1987}   & \multicolumn{1}{c}{0.2608}  & 0.8000      & \multicolumn{1}{c}{0.4800}      & \multicolumn{1}{c}{0.0385}   & \multicolumn{1}{c}{0.2512} & 0.7600      \\ 
PCNN       & \multicolumn{1}{c}{0.4400}      & \multicolumn{1}{c}{0.1143}   & \multicolumn{1}{c}{0.2568}  & 0.7800      & \multicolumn{1}{c}{0.4800}      & \multicolumn{1}{c}{0.0385}   & \multicolumn{1}{c}{0.2512} & 0.7600      \\ \hline
\end{tabular}
}
\label{tab:glcm_pred_pattern}
\end{table}

\subsubsection{Case Study -- shifting 1s vector in all 0s matrix horizontally or vertically}\label{sec:pat_case_2}
Consider a weave pattern with all 0s. Then we artificially change each row of the matrix to be 1s. When all 1s row lies between the top and bottom row of the pattern, the corresponding GLCMs are the same. When all 1s row lies on top or bottom of the pattern, the GLCM will be different from others. However, for our woven analysis, we are picking the RUC of the model and using periodic boundary conditions, which means viewing globally, the GLCM for the woven pattern with all 1's rows at the bottom or top will be the same as in other places. Thus, GLCM will be the same no matter where we put them all 1's row. Furthermore, we prove from the FEA result that the woven modulus is not changing for all 1's rows at different positions. When we further shift all 1s vectors at different columns of the all 0s pattern, where GLCM is also the same, we notice that the modulus is not changing. \emph{These results tell us that when GLCM are close, the two woven models' modulus will also be close. That is to say, GLCM controls the features that determine the woven model's modulus}.

\subsubsection{Case Study -- weave pattern optimization through GLCM}\label{sec:pat_case_3}

\begin{figure}[h!]
\centering
\subfigure[]{
  \includegraphics[width=0.3\textwidth]{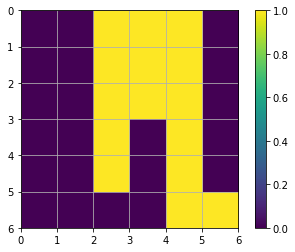}
}
\subfigure[]{
  \includegraphics[width=0.3\textwidth]{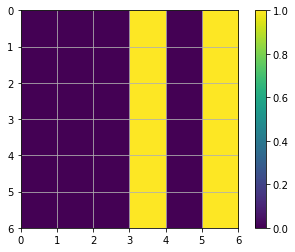}
}
\caption{woven composite patterns for Case 3}
\label{img:case_3}
\end{figure}

We start by considering a weave pattern predicted by the PCNN-BDPa shown in Figure~\ref{img:case_3}(a), which meets our desired mechanical properties requirement. The four corresponding GLCMs are: 

\begin{equation}
\begin{bmatrix}
    16 & 15 \\
    15 & 14
\end{bmatrix} \quad
\begin{bmatrix}
    32 & 3 \\
    3 & 32
\end{bmatrix} \quad
\begin{bmatrix}
    12 & 13 \\
    13 & 12
\end{bmatrix} \quad
\begin{bmatrix}
    12 & 14 \\
    14 & 10
\end{bmatrix}
\end{equation}

If we want to optimize the prediction to achieve an even higher modulus, we can modify the GLCM. From our previous lemmas, we know that we have to achieve higher energy and homogeneity in GLCM. Thus, we can change the GLCM in a horizontal direction such that some elements in the GLCM are high and others are very low. Then, we can convert it back to a weave pattern and adjust the pattern to increase energy in other GLCMs. An example of a modified pattern is shown in Figure~\ref{img:case_3}(b), whose GLCM are:

\begin{equation}
\begin{bmatrix}
    24 & 18 \\
    18 & 0
\end{bmatrix} \quad
\begin{bmatrix}
    40 & 0 \\
    0 & 20
\end{bmatrix} \quad
\begin{bmatrix}
    20 & 15 \\
    15 & 0
\end{bmatrix} \quad
\begin{bmatrix}
    20 & 15 \\
    15 & 0
\end{bmatrix}
\end{equation}

Here, we can easily tell that the second woven pattern has higher energy than the first. By further evaluation in FEA, we notice that if we use this pattern for a single material woven composite, the original pattern has modulus: $E_1=44.3$ GPa, $E_2=26.3$ GPa, and the modified pattern has modulus $E_1=49.5$ GPa, $E_2=26.5$ GPa. The improvement is more significant when considering bi-material woven composite, as each modulus's variations are more significant. Here we will show one case where we assume the assigned material is an all '1's vector, and through FEA, we determine that the original woven pattern has a modulus of $E_1=28.4$ GPa, $E_2=16.5$ GPa, $G_{12}=1.96$ GPa and the modified woven pattern has a modulus $E_1=35.2$ GPa, $E_2=20.2$ GPa, $G_{12}=2.30$ GPa. \emph{Thus, GLCM gives us a way to validate the proposed weave pattern by PCNN and further enhance the modulus if needed}.

\textit{Analytical validation:} To further validate the GLCM optimization, we will compare both models qualitatively using the weave parameters discussed in Section~\ref{app:WeavePar}. Comparing both weave architectures in Figure \ref{img:case_3}, we observe that GLCM optimization has removed and reduced the regions of undulations in warp and weft directions, respectively. In previous research \cite{ishikawa1982,osada2003initial}, it has been shown that the regions of undulation result in lower in-plane moduli values. Therefore, the GLCM-optimized pattern has higher in-plane moduli than the weave pattern obtained from the PCNN-BDPa. 

\subsection{Weave Material Optimization}
\begin{figure}[h!]
\centering
\subfigure[]{
  \includegraphics[width=0.3\textwidth]{./figures/case_3_model_1.png}
}
\subfigure[]{
    \includegraphics[width=0.3\textwidth]{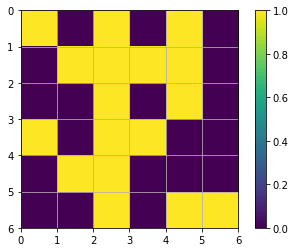}
}
\caption{Two weave patterns considered for weave material sequence case studies}
\label{img:woven_mat_study_pat}
\end{figure}

In this section, we further show weave material sequence optimization by conducting case studies of 2 randomly chosen patterns shown in Figure~\ref{img:woven_mat_study_pat}. For each pattern, different choices of material vectors are considered. 

\subsubsection{Case Study -- weave material sequence sensitivity study} \label{sec:case_mat_1}

First, we consider two 1-by-6 material vectors, where the first material vector represents horizontal yarn materials and the second represents vertical yarn materials. We consider two scenarios: (1) The ratio between the two materials in each vector is $1:1$. (2) the ratio between the two materials in each material vector is $3:1$ and $1:3$, respectively. Then, we randomly choose several woven patterns, for example, as shown in Figure~\ref{img:woven_mat_study_pat}. Regardless of our material vector sequence, the woven composite's in-plane modulus will be nearly the same if the ratio between the two materials is the same for horizontal and vertical yarns. \emph{This validates our conclusion that only the material vector's mean or sum controls the corresponding in-plane modulus.}

\subsubsection{Case Study -- weave material sequence optimization for specific weave patterns with different material ratios} \label{sec:case_mat_2}

\paragraph{Overall material ratio 1:1}
Assume initially, we either have a material vector $M_b$ or is proposed by PCNN as:

\begin{equation}
M_b = \begin{bmatrix}
    1 & 0 & 1 & 1 & 0 & 1 \\
    0 & 1 & 0 & 0 & 0 & 1
\end{bmatrix}
\end{equation}

The first row represents the horizontal yarn materials, and the second represents the vertical yarn materials. Such material assignment gives us the corresponding modulus $E_1=40.18$ GPa, $E_2=23.39$ GPa, $G_{12}=2.702$ GPa. Now, we want to enhance the corresponding overall modulus of the woven pattern. Here, we introduce two material vectors; one completely follows our conclusion from the regression-based analysis in Section~\ref{sec:mat_feature_ana} (say $M_c$), and the other reverses our conclusion (say $M_r$). The corresponding material vectors can be expressed as:

\begin{equation}
M_c =\begin{bmatrix}
    0 & 0 & 0 & 0 & 0 & 0 \\
    1 & 1 & 1 & 1 & 1 & 1
\end{bmatrix}
 \quad  \mathrm{and} \quad M_r=
\begin{bmatrix}
    1 & 1 & 1 & 1 & 1 & 1 \\
    0 & 0 & 0 & 0 & 0 & 0
\end{bmatrix}
\end{equation}

\paragraph{Overall material ratio 1:3}
Here, our material vector is:
\begin{equation}
M_b=
\begin{bmatrix}
    1 & 1 & 0 & 1 & 0 & 1 \\
    1 & 1 & 0 & 1 & 1 & 1
\end{bmatrix}
\end{equation}

This material assignment gives us a modulus of $E_1=40.18$ GPa, $E_2=23.39$ GPa, $G_{12}=2.702$ GPa. To enhance the corresponding overall modulus of the woven pattern, we introduce two material vectors as above:
\begin{equation}
M_c=
\begin{bmatrix}
    0 & 0 & 0 & 1 & 1 & 1 \\
    1 & 1 & 1 & 1 & 1 & 1
\end{bmatrix}
  \quad  \mathrm{and} \quad M_r=
\begin{bmatrix}
    1 & 1 & 1 & 1 & 1 & 1 \\
    1 & 1 & 1 & 0 & 0 & 0
\end{bmatrix}
\end{equation}

\paragraph{Overall material ratio 3:1}
For this case study, we have a material vector: 
\begin{equation}
M_b=
\begin{bmatrix}
    1 & 0 & 0 & 1 & 0 & 0 \\
    0 & 0 & 0 & 1 & 0 & 0
\end{bmatrix}
\end{equation}

The corresponding modulus are $E_1=40.18$ GPa, $E_2=23.39$ GPa, and $G_{12}=2.702$ GPa. We introduce the two material vectors $M_c$ and $M_r$ as:

\begin{equation}
M_c=
\begin{bmatrix}
    0 & 0 & 0 & 0 & 0 & 0 \\
    0 & 0 & 0 & 1 & 1 & 1
\end{bmatrix}
  \quad  \mathrm{and} \quad M_r=
\begin{bmatrix}
    1 & 1 & 1 & 0 & 0 & 0 \\
    0 & 0 & 0 & 0 & 0 & 0
\end{bmatrix}
\end{equation}

\paragraph{Results}

From Table~\ref{tab:mat_case_study}, the sequence $M_c$, which follows our optimization strategy, consistently achieves superior overall modulus over $M_b$ and $M_r$. While, $M_r$, which contradicts our optimization strategy, nearly always achieves the lowest modulus. \emph{Thus, using the GLCM based optimization strategy, we can conclude that having more material 1 in the vertical yarn and material 0 in horizontal yarn is beneficial for the patterns shown in Figure \ref{img:woven_mat_study_pat}}. This can be extended to other patterns considered by a user. 

\begin{table}[h!]
\centering
\caption{Weave material sequence based optimization}
\resizebox{0.99\textwidth}{!}{
\begin{tabular}{lccccccccccccc}
\hline
                           & Material Ratio & \multicolumn{4}{c}{Ratio 1:1}                                                                                    & \multicolumn{4}{c}{Ratio 1:3}                                                                                    & \multicolumn{4}{c}{Ratio 3:1}                                                                                    \\ 
                           & Modulus         & \multicolumn{1}{c}{E1-N}    & \multicolumn{1}{c}{E2-N}    & \multicolumn{1}{c}{G12-N}   & Sum                        & \multicolumn{1}{c}{E1-N}    & \multicolumn{1}{c}{E2-N}    & \multicolumn{1}{c}{G12-N}   & Sum                        & \multicolumn{1}{c}{E1-N}    & \multicolumn{1}{c}{E2-N}    & \multicolumn{1}{c}{G12-N}   & Sum                        \\ \hline
\multirow{3}{*}{Pattern 1} & $M_b$ (GPa)      & \multicolumn{1}{l}{32.65} & \multicolumn{1}{l}{20.76} & \multicolumn{1}{l}{23.93} & \multicolumn{1}{l}{77.34}                      & \multicolumn{1}{l}{31.15} & \multicolumn{1}{l}{17.95} & \multicolumn{1}{l}{21.61} & \multicolumn{1}{l}{70.72}                      & \multicolumn{1}{l}{36.80} & \multicolumn{1}{l}{23.32} & \multicolumn{1}{l}{27.11} & \multicolumn{1}{l}{87.23}                      \\ 
                           & $M_c$ (GPa)       & \multicolumn{1}{l}{36.33} & \multicolumn{1}{l}{20.03} & \multicolumn{1}{l}{25.04} & \multicolumn{1}{l}{81.40}                      & \multicolumn{1}{l}{32.23} & \multicolumn{1}{l}{17.81} & \multicolumn{1}{l}{21.98} & \multicolumn{1}{l}{72.01}                      & \multicolumn{1}{l}{38.70} & \multicolumn{1}{l}{22.57} & \multicolumn{1}{l}{27.60} & \multicolumn{1}{l}{88.86}                      \\ 
                           & $M_r$ (GPa)       & \multicolumn{1}{l}{31.35} & \multicolumn{1}{l}{21.50} & \multicolumn{1}{l}{23.87} & \multicolumn{1}{l}{76.72}                      & \multicolumn{1}{l}{29.39} & \multicolumn{1}{l}{18.89} & \multicolumn{1}{l}{21.50} & \multicolumn{1}{l}{69.78}                      & \multicolumn{1}{l}{36.33} & \multicolumn{1}{l}{24.02} & \multicolumn{1}{l}{27.32} & \multicolumn{1}{l}{87.67}                      \\ \hline
\multirow{3}{*}{Pattern 2} & $M_b$ (GPa)       & \multicolumn{1}{c}{27.94} & \multicolumn{1}{l}{24.51} & \multicolumn{1}{l}{23.83} & \multicolumn{1}{l}{76.27} & \multicolumn{1}{l}{26.51} & \multicolumn{1}{l}{20.81} & \multicolumn{1}{l}{21.23} & \multicolumn{1}{l}{68.55} & \multicolumn{1}{l}{31.92} & \multicolumn{1}{l}{27.20} & \multicolumn{1}{l}{26.93} & \multicolumn{1}{l}{86.04} \\  
                           & $M_c$ (GPa)       & \multicolumn{1}{l}{31.67} & \multicolumn{1}{l}{23.04} & \multicolumn{1}{l}{24.51} & \multicolumn{1}{l}{79.23} & \multicolumn{1}{l}{27.69} & \multicolumn{1}{l}{20.51} & \multicolumn{1}{l}{21.46} & \multicolumn{1}{l}{69.56} & \multicolumn{1}{l}{34.05} & \multicolumn{1}{l}{26.12} & \multicolumn{1}{l}{27.19} & \multicolumn{1}{l}{87.35} \\ 
                           & $M_r$ (GPa)       & \multicolumn{1}{l}{26.47} & \multicolumn{1}{l}{25.67} & \multicolumn{1}{l}{23.87} & \multicolumn{1}{l}{76.01} & \multicolumn{1}{l}{24.76} & \multicolumn{1}{l}{22.42} & \multicolumn{1}{l}{21.34} & \multicolumn{1}{l}{68.52} & \multicolumn{1}{l}{31.49} & \multicolumn{1}{l}{28.40} & \multicolumn{1}{l}{27.38} & \multicolumn{1}{l}{87.28} \\ \hline
\end{tabular}
}
\label{tab:mat_case_study}
\end{table}

\subsection{Summary of Regression and Optimization Results}

Based on the regression analysis and optimization studies conducted on the weave pattern and material sequence, we conclude that:
\begin{enumerate}
    \item When optimizing the weave pattern with a fixed material sequence, the relationship between the weave pattern and overall in-plane modulus can be effectively described using GLCM features. The regression analysis reveals that energy and homogeneity positively correlate with the overall modulus, while contrast and correlation exhibit a negative correlation. This conclusion holds for all checkerboard-type models and can be utilized to optimize the weave pattern.
    \item Compared to the GIDN framework with different random samples, our proposed GLCM-based optimization approach outperforms GIDNs by achieving optimal designs and requiring less time, as it does not rely on random searching.
    \item When considering a fixed weave pattern, the relationship between the weave material sequence and overall in-plane modulus can be described by examining the mean of the material sequence vector in the physical space. It is important to note that this conclusion is specific to patterns in Figure\ref{img:woven_mat_study_pat} and two materials considered in the study.
\end{enumerate}                                                                                                                            

\subsection{Discover Optimal Woven Composite Architecture at Initial Design Stage}

We have shown that PCNN can predict weave patterns or material sequences with high accuracy, and feature-based optimization can enhance the overall modulus of woven composite models. The proposed optimization strategy can be combined with PCNN to determine the optimal woven composite architecture even at the initial design stage. For example, to design a woven composite model, we assume there are two materials to choose from for each yarn. We can use any weave pattern to find the woven composite with the highest overall modulus. To determine the optimal design, we can follow the steps below:
\begin{enumerate}
    \item Determine the optimal material sequence through the feature-based optimization strategy, then follow the procedure described in Section~\ref{sec:mat_stat_feature} and Section~\ref{sec:regress_feature}.
    \item Pick the maximum in-plane modulus ($E_1$, $E_2$, $G_{12}$) within a reasonable range, choose the material sequence vector determined in Step 1, then use PCNN to predict the weave pattern. 
    \item After obtaining the weave pattern, further utilize a feature-based optimization strategy to optimize the weave pattern to achieve the optimal woven composite designs.
\end{enumerate}

\section{Conclusions}

The objective of this paper was twofold: 1) The first was to establish a bridge between woven architectures (patterns and material sequences) and the corresponding in-plane modulus by developing Deep Neural Networks. We classify the prediction into the typical design process (FDP) and inverse design process (BDPa and BDPb). The FDP problem is solved by a Deep Convolutional Neural Network (DCNN). For the much more complex BDP problems, we proposed the Physics-Constraint Neural Network (PCNN) to predict the woven composite architecture from the in-plane modulus. We have shown that our proposed DCNN delivers relatively accurate predictions. More importantly, PCNN can make sound predictions for BDP problems and vastly outperforms the baseline models we considered. 2) The second was to propose a feature-based optimization strategy to find optimal woven composite architectures. We proposed a GLCM feature-based optimization strategy for weave patterns and statistical feature-based optimization for weave material sequences. We further showed that the feature-based optimization strategy can accurately and conveniently optimize the woven composite architecture. Finally, we showed how to find the optimal woven composite architecture by combining PCNN with our proposed feature-based optimization strategy.

Key contributions to this paper are:
\begin{enumerate}
    \item To our knowledge, this is the first attempt toward a bi-directional design process for woven fabrics and textiles with Deep Neural Networks. That is, predicting mechanical properties from weave architectures (pattern and material sequence) and vice-versa. We primarily focused on woven composites in this paper. However, this approach can be applied to generic woven fabrics and textiles.
    \item To solve the complex backward prediction (BDP) problems, we proposed our Physics-Constraint Neural Network (PCNN) to bridge the woven composite's modulus and architecture. We have shown that our proposed Neural Network vastly increases the prediction accuracy compared to several well-established baseline models.
    \item We further proposed feature-based optimization to optimize the woven composite architecture. We proposed a Gray Level Co-occurrence Matrix-based optimization strategy for weave pattern optimization and a statistical feature-based optimization strategy for weave material sequence. The feature-based optimization strategy can be combined with PCNN to determine the optimal woven composite architecture even at the initial design stage.

\end{enumerate}

\section*{Credit Authorship Contribution Statement}
\textbf{H. F.} `contributed' Conceptualization, Methodology, Validation, Formal Analysis, Investigation, Writing the manuscript.
\\
\textbf{S. P. S.} `contributed' Methodology, Validation, Formal analysis, Investigation, and Writing the manuscript. 
\\
\textbf{H. R. T.} `contributed' Validation, Formal analysis, Investigation, and Writing the manuscript. 
\\
\textbf{P. P.} `contributed' Conceptualization, Validation, Formal analysis, Investigation, Writing the manuscript, Project management, and Funding acquisition.

\section*{Declaration of Competing Interest}
The authors declare that they have no known competing financial interests or personal relationships that could have appeared to influence the work reported in this paper.

\section*{Materials and Correspondence}
All correspondence and material requests should be addressed to the corresponding author - Pavana Prabhakar at pavana.prabhakar@wisc.edu.


\section*{Acknowledgements}
The authors would like to acknowledge the support through the NSF CAREER award through {\em{Mechanics of Materials and Structures (MOMS) Program}} for conducting the research presented here. The authors would also like to acknowledge the support from the University of Wisconsin Graduate Fellowship.


\section*{Data Availability}

The data that support the findings of this study are available from the corresponding author upon reasonable request.




{\footnotesize
\bibliographystyle{unsrt}
\bibliography{sample}

\begin{thebibliography}{10}

\bibitem{mouritz1999review}
AP~Mouritz, MK~Bannister, PJ~Falzon, and KH~Leong.
\newblock Review of applications for advanced three-dimensional fibre textile
  composites.
\newblock {\em Composites Part A: applied science and manufacturing},
  30(12):1445--1461, 1999.

\bibitem{kelkar2006structural}
AD~Kelkar, JS~Tate, and R~Bolick.
\newblock Structural integrity of aerospace textile composites under fatigue
  loading.
\newblock {\em Materials Science and Engineering: B}, 132(1-2):79--84, 2006.

\bibitem{carey2017introduction}
JP~Carey, GW~Melenka, AJ~Hunt, and C~Ayranci.
\newblock Introduction to braided composite material behavior.
\newblock In {\em Handbook of Advances in Braided Composite Materials}, pages
  207--237. Elsevier, 2017.

\bibitem{gao2021textile}
Y~Gao, C~Xie, and Z~Zheng.
\newblock Textile composite electrodes for flexible batteries and
  supercapacitors: opportunities and challenges.
\newblock {\em Advanced Energy Materials}, 11(3):2002838, 2021.

\bibitem{long2005design}
AC~Long.
\newblock {\em Design and manufacture of textile composites}.
\newblock Elsevier, 2005.

\bibitem{naik1992prediction}
NK~Naik and VK~Ganesh.
\newblock Prediction of on-axes elastic properties of plain weave fabric
  composites.
\newblock {\em Composites Science and Technology}, 45(2):135--152, 1992.

\bibitem{jiang2006investigation}
YP~Jiang, WL~Guo, and ZF~Yue.
\newblock Investigation of the three-dimensional micromechanical behavior of
  woven-fabric composites.
\newblock {\em Mechanics of Composite Materials}, 42(2):141--150, 2006.

\bibitem{khan2017finite}
HA~Khan, A~Hassan, MB~Saeed, F~Mazhar, and IA~Chaudhary.
\newblock Finite element analysis of mechanical properties of woven composites
  through a micromechanics model.
\newblock {\em Science and Engineering of Composite Materials}, 24(1):87--99,
  2017.

\bibitem{ishikawa1983one}
T~Ishikawa and TW~Chou.
\newblock One-dimensional micromechanical analysis of woven fabric composites.
\newblock {\em AIAA journal}, 21(12):1714--1721, 1983.

\bibitem{whitcomb1991three}
JD~Whitcomb.
\newblock Three-dimensional stress analysis of plain weave composites.
\newblock {\em Composite materials: Fatigue and fracture.}, 3:417--438, 1991.

\bibitem{whitcomb1994macro}
J~Whitcomb, K~Woo, and S~Gundapaneni.
\newblock Macro finite element for analysis of textile composites.
\newblock {\em Journal of composite materials}, 28(7):607--618, 1994.

\bibitem{whitcomb1995boundary}
J~Whitcomb, G~Kondagunta, and K~Woo.
\newblock Boundary effects in woven composites.
\newblock {\em Journal of composite materials}, 29(4):507--524, 1995.

\bibitem{gowayed2013types}
Y~Gowayed.
\newblock Types of fiber and fiber arrangement in fiber-reinforced polymer
  (frp) composites.
\newblock In {\em Developments in Fiber-Reinforced Polymer (FRP) Composites for
  Civil Engineering}, pages 3--17. Elsevier, 2013.

\bibitem{dong2016experimental}
K~Dong, K~Liu, Q~Zhang, B~Gu, and B~Sun.
\newblock Experimental and numerical analyses on the thermal conductive
  behaviors of carbon fiber/epoxy plain woven composites.
\newblock {\em International Journal of Heat and Mass Transfer}, 102:501--517,
  2016.

\bibitem{krizhevsky2012imagenet}
A~Krizhevsky, I~Sutskever, and GE~Hinton.
\newblock Imagenet classification with deep convolutional neural networks.
\newblock {\em Advances in neural information processing systems}, 25, 2012.

\bibitem{creswell2018generative}
A~Creswell, T~White, V~Dumoulin, K~Arulkumaran, B~Sengupta, and AA~Bharath.
\newblock Generative adversarial networks: An overview.
\newblock {\em IEEE Signal Processing Magazine}, 35(1):53--65, 2018.

\bibitem{ying2018graph}
R~Ying, R~He, K~Chen, P~Eksombatchai, WL~Hamilton, and J~Leskovec.
\newblock Graph convolutional neural networks for web-scale recommender
  systems.
\newblock In {\em Proceedings of the 24th ACM SIGKDD international conference
  on knowledge discovery \& data mining}, pages 974--983, 2018.

\bibitem{zhang2015deep}
W~Zhang, R~Li, H~Deng, L~Wang, W~Lin, S~Ji, and D~Shen.
\newblock Deep convolutional neural networks for multi-modality isointense
  infant brain image segmentation.
\newblock {\em NeuroImage}, 108:214--224, 2015.

\bibitem{conneau2016very}
A~Conneau, H~Schwenk, L~Barrault, and Y~Lecun.
\newblock Very deep convolutional networks for natural language processing.
\newblock {\em arXiv preprint arXiv:1606.01781}, 2(1), 2016.

\bibitem{schlegl2017unsupervised}
T~Schlegl, P~Seeb{\"o}ck, SM~Waldstein, U~Schmidt-Erfurth, and G~Langs.
\newblock Unsupervised anomaly detection with generative adversarial networks
  to guide marker discovery.
\newblock In {\em International conference on information processing in medical
  imaging}, pages 146--157. Springer, 2017.

\bibitem{bousmalis2017unsupervised}
K~Bousmalis, N~Silberman, D~Dohan, D~Erhan, and D~Krishnan.
\newblock Unsupervised pixel-level domain adaptation with generative
  adversarial networks.
\newblock In {\em Proceedings of the IEEE conference on computer vision and
  pattern recognition}, pages 3722--3731, 2017.

\bibitem{souly2017semi}
N~Souly, C~Spampinato, and M~Shah.
\newblock Semi supervised semantic segmentation using generative adversarial
  network.
\newblock In {\em Proceedings of the IEEE international conference on computer
  vision}, pages 5688--5696, 2017.

\bibitem{zhao2016energy}
J~Zhao, M~Mathieu, and Y~LeCun.
\newblock Energy-based generative adversarial network.
\newblock {\em arXiv preprint arXiv:1609.03126}, 2016.

\bibitem{yu2017seqgan}
L~Yu, W~Zhang, J~Wang, and Y~Yu.
\newblock Seqgan: Sequence generative adversarial nets with policy gradient.
\newblock In {\em Proceedings of the AAAI conference on artificial
  intelligence}, volume~31, 2017.

\bibitem{wei2018predicting}
H~Wei, S~Zhao, Q~Rong, and H~Bao.
\newblock Predicting the effective thermal conductivities of composite
  materials and porous media by machine learning methods.
\newblock {\em International Journal of Heat and Mass Transfer}, 127:908--916,
  2018.

\bibitem{chen2019machine}
CT~Chen and GX~Gu.
\newblock Machine learning for composite materials.
\newblock {\em MRS Communications}, 9(2):556--566, 2019.

\bibitem{feng2021difference}
H~Feng and P~Prabhakar.
\newblock Difference-based deep learning framework for stress predictions in
  heterogeneous media.
\newblock {\em Composite Structures}, 269:113957, 2021.

\bibitem{bang2020defect}
HT~Bang, S~Park, and H~Jeon.
\newblock Defect identification in composite materials via thermography and
  deep learning techniques.
\newblock {\em Composite Structures}, 246:112405, 2020.

\bibitem{liu2019initial}
X~Liu, F~Gasco, J~Goodsell, and W~Yu.
\newblock Initial failure strength prediction of woven composites using a new
  yarn failure criterion constructed by deep learning.
\newblock {\em Composite Structures}, 230:111505, 2019.

\bibitem{nardi2021design}
D~Nardi and J~Sinke.
\newblock Design analysis for thermoforming of thermoplastic composites:
  Prediction and machine learning-based optimization.
\newblock {\em Composites Part C: Open Access}, 5:100126, 2021.

\bibitem{sepasdar2021data}
R~Sepasdar, A~Karpatne, and M~Shakiba.
\newblock A data-driven approach to full-field damage and failure pattern
  prediction in microstructure-dependent composites using deep learning.
\newblock {\em arXiv preprint arXiv:2104.04485}, 2021.

\bibitem{gu2018bioinspired}
GX~Gu, CT~Chen, DJ~Richmond, and MJ~Buehler.
\newblock Bioinspired hierarchical composite design using machine learning:
  simulation, additive manufacturing, and experiment.
\newblock {\em Materials Horizons}, 5(5):939--945, 2018.

\bibitem{abueidda2019prediction}
DW~Abueidda, M~Almasri, R~Ammourah, U~Ravaioli, IM~Jasiuk, and NA~Sobh.
\newblock Prediction and optimization of mechanical properties of composites
  using convolutional neural networks.
\newblock {\em Composite Structures}, 227:111264, 2019.

\bibitem{feng2021deep}
H~Feng, SP~Subramaniyan, and P~Prabhakar.
\newblock Deep learning framework for woven composite analysis.
\newblock In {\em Proceedings of the American Society for
  Composites—Thirty-Sixth Technical Conference on Composite Materials}, 2021.

\bibitem{chen2020generative}
CT~Chen and GX~Gu.
\newblock Generative deep neural networks for inverse materials design using
  backpropagation and active learning.
\newblock {\em Advanced Science}, 7(5):1902607, 2020.

\bibitem{mao2020physics}
Z~Mao, AD~Jagtap, and GE~Karniadakis.
\newblock Physics-informed neural networks for high-speed flows.
\newblock {\em Computer Methods in Applied Mechanics and Engineering},
  360:112789, 2020.

\bibitem{pan2009survey}
SJ~Pan and Q~Yang.
\newblock A survey on transfer learning.
\newblock {\em IEEE Transactions on knowledge and data engineering},
  22(10):1345--1359, 2009.

\bibitem{weiss2016survey}
K~Weiss, TM~Khoshgoftaar, and D~Wang.
\newblock A survey of transfer learning.
\newblock {\em Journal of Big data}, 3(1):1--40, 2016.

\bibitem{badrinarayanan2017segnet}
V~Badrinarayanan, A~Kendall, and R~Cipolla.
\newblock Segnet: A deep convolutional encoder-decoder architecture for image
  segmentation.
\newblock {\em IEEE transactions on pattern analysis and machine intelligence},
  39(12):2481--2495, 2017.

\bibitem{sebastian2012gray}
B~Sebastian~V, A~Unnikrishnan, and K~Balakrishnan.
\newblock Gray level co-occurrence matrices: generalisation and some new
  features.
\newblock {\em arXiv preprint arXiv:1205.4831}, 2012.

\bibitem{zulpe2012glcm}
N~Zulpe and V~Pawar.
\newblock Glcm textural features for brain tumor classification.
\newblock {\em International Journal of Computer Science Issues (IJCSI)},
  9(3):354, 2012.

\bibitem{singh2012classification}
D~Singh and K~Kaur.
\newblock Classification of abnormalities in brain mri images using glcm, pca
  and svm.
\newblock {\em International Journal of Engineering and Advanced Technology
  (IJEAT)}, 1(6):243--248, 2012.

\bibitem{hall2017glcm}
M~Hall-Beyer.
\newblock Glcm texture: A tutorial v. 3.0 march 2017.
\newblock 2017.

\bibitem{raheja2013fabric}
JL~Raheja, S~Kumar, and A~Chaudhary.
\newblock Fabric defect detection based on glcm and gabor filter: A comparison.
\newblock {\em Optik}, 124(23):6469--6474, 2013.

\bibitem{haralick1973textural}
RM~Haralick, K~Shanmugam, and IH~Dinstein.
\newblock Textural features for image classification.
\newblock {\em IEEE Transactions on systems, man, and cybernetics},
  (6):610--621, 1973.

\bibitem{ABAQUS}
ABAQUS (2011)~Dassualt Systèmes.
\newblock Abaqus documentation.
\newblock 2011.

\bibitem{Lin2011}
H~Lin, LP~Brown, and AC~Long.
\newblock {Modelling and simulating textile structures using TexGen}.
\newblock {\em Advanced Materials Research}, 331:44--47, 2011.

\bibitem{Chamis1989}
CC~Chamis.
\newblock {Mechanics of composite materials: Past, present, and future}.
\newblock {\em Journal of Composites Technology and Research}, 11(1):3--14,
  1989.

\bibitem{Li2004}
S~Li and A~Wongsto.
\newblock {Unit cells for micromechanical analyses of particle-reinforced
  composites}.
\newblock {\em Mechanics of Materials}, 36(7):543--572, 2004.

\bibitem{ishikawa1982}
T~Ishikawa and T~W Chou.
\newblock Stiffness and strength behaviour of woven fabric composites.
\newblock {\em Journal of Materials Science}, 17:3211--3220, 1982.

\bibitem{pang2021dislocation}
GD~Pang, YC~Lin, YL~Qiu, YQ~Jiang, YW~Xiao, and MS~Chen.
\newblock Dislocation density--based model and stacked auto-encoder model for
  ti-55511 alloy with basket-weave microstructures deformed in $\alpha$+
  $\beta$ region.
\newblock {\em Advanced Engineering Materials}, 23(4):2001307, 2021.

\bibitem{jiang2021stressgan}
H~Jiang, Z~Nie, R~Yeo, AB~Farimani, and LB~Kara.
\newblock Stressgan: A generative deep learning model for two-dimensional
  stress distribution prediction.
\newblock {\em Journal of Applied Mechanics}, 88(5), 2021.

\bibitem{whitley1994genetic}
D~Whitley.
\newblock A genetic algorithm tutorial.
\newblock {\em Statistics and computing}, 4(2):65--85, 1994.

\bibitem{haralick1992computer}
RM~Haralick and LG~Shapiro.
\newblock {\em Computer and robot vision}, volume~1.
\newblock Addison-wesley Reading, 1992.

\bibitem{bevk2002statistical}
M~Bevk and I~Kononenko.
\newblock A statistical approach to texture description of medical images: a
  preliminary study.
\newblock In {\em Proceedings of 15th IEEE Symposium on Computer-Based Medical
  Systems (CBMS 2002)}, pages 239--244. IEEE, 2002.

\bibitem{osada2003initial}
T~Osada, A~Nakai, and H~Hamada.
\newblock Initial fracture behavior of satin woven fabric composites.
\newblock {\em Composite structures}, 61(4):333--339, 2003.

\end{thebibliography}
}


\newpage

\section*{Supplementary Document}
\addcontentsline{toc}{section}{Appendices}
\renewcommand{\thesubsection}{\Alph{subsection}}
\renewcommand{\thefigure}{\Alph{subsection}.\arabic{figure}}
\setcounter{figure}{0}

\subsection{Explaining the weave parameters}\label{app:WeavePar}

\subsubsection{$n_{g}$ and Crimp ratio}

\begin{figure}[h!]
\centering
	\includegraphics[width=0.9\textwidth]{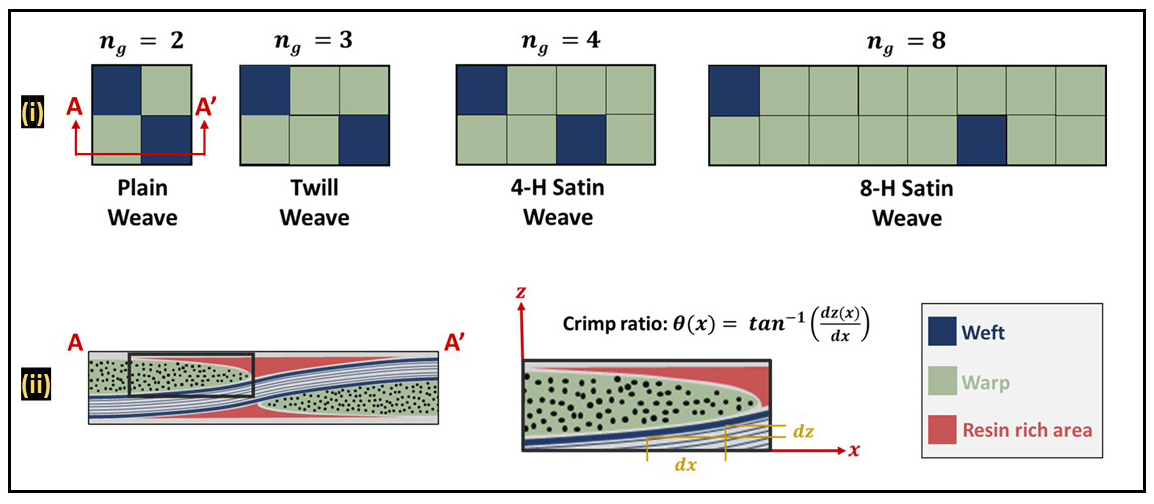}
\caption{(i) Examples of $n_{g}$ values for Plain, Twill, 4-H Satin, and 8-H Satin weave patterns and (ii) Section AA' of the plain weave to illustrate crimp ratio ($\theta(x)$) at the interlaced region.}
\label{img:Factor1}
\end{figure}

Woven fabrics comprise sets of warp and weft threads that are interlaced together in different ways to achieve various architectures. Earlier, the weave patterns consisted of uniform interlacing of these yarns in perpendicular directions, and these patterns could be classified using the repeat of the interlaced regions. A geometrical parameter, $n_{g}$ defines the number of warp yarns that are interlaced with one single weft yarn. In Figure~\ref{img:Factor1}(i), we have shown traditional weave patterns with their respective $n_{g}$ values. As we can observe a plain weave has the lowest $n_{g}$ value of 2 whereas the 8-harness (8-H) satin has an $n_{g}$ value of 8. Another geometrical factor, crimp ratio ($\theta_{g}$), reflects the undulation of the yarns at the interlaced region as shown in Figure~\ref{img:Factor1}(ii). For 2D woven structures, the crimp ratio is defined for warp and weft directions. For a woven fabric, the crimp ratio increases with a decrease in $n_{g}$ value and vice-versa. For example, the crimp ratio of a plain weave will be higher than that of a plain weave. Osada et al. \cite{osada2003initial} compared the failure of composites with plain weave and 5-H satin weave, and they reported that the crimp ratio had a significant impact on the material properties. It was shown that the plain weave had a crimp ratio of 0.164 whereas the value was 0.023 for the satin weave. They also proposed that the initial slope and the strength of the satin weave composite were higher than the plain weave composite, with a delayed knee-point formation. Ishikawa and Chou \cite{ishikawa1982} also exploited these geometrical parameters to propose an analytical model to predict the properties of woven composites. The model developed showed that the elastic moduli of composites reduce with the existence of the undulated regions in the warp and weft directions, respectively. 

\subsubsection{Bridging effect}
\begin{figure}[h!]
\centering
	\includegraphics[width=0.75\textwidth]{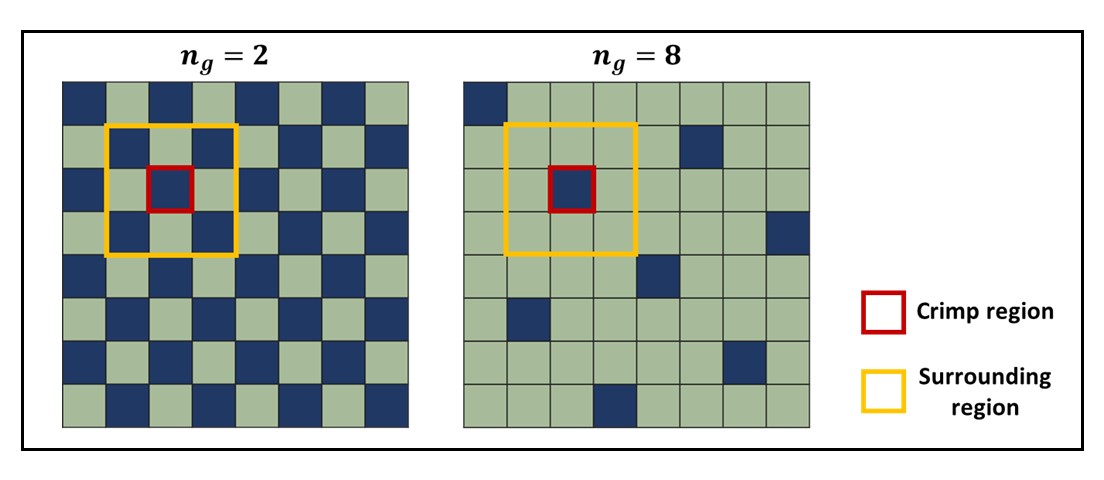}
\caption{Illustration of bridging concept for Plain and 8-H weave patterns. The red box depicts the area of undulation, while the yellow box depicts the surrounding region.}
\label{img:Factor2}
\end{figure}

Ishikawa and Chou \cite{ishikawa1982} also proposed a "bridging" model to predict the mechanical properties of satin composites ($n_{g} \geq 4$). In this model, the surrounding region around the crimp region was considered to obtain the properties. In Figure~\ref{img:Factor2}, we have shown a comparison between plain weave and 8-H satin weaves. In plain weave, we can observe that the surrounding regions also consist of the undulated region which will result in lower in-plane moduli. On the other hand, for 8-H satin weave, the surrounding region consists of straight yarns with no undulated region. Therefore, the surrounding region has higher local in-plane moduli compared to the undulated region. These straight yarns in surrounding regions act as a load-carrying bridge between neighboring interlaced regions. 

Although the previous research is restricted to the mechanical properties of uniform weave architectures, it has been shown that geometrical parameters of a weave architecture play a vital role in determining the mechanical behavior of woven composites. In this work, we will utilize these geometric parameters to evaluate the predictions made using the Machine Learning framework.

\subsection{Woven composite models having same in-plane modulus}\label{App:Woven_Hist}
\begin{figure}[h!]
\centering
\subfigure[]{
  \includegraphics[width=0.3\textwidth]{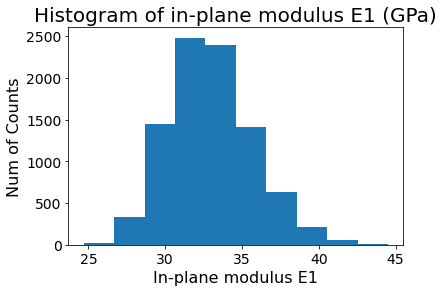}
}
\centering
\subfigure[]{
  \includegraphics[width=0.3\textwidth]{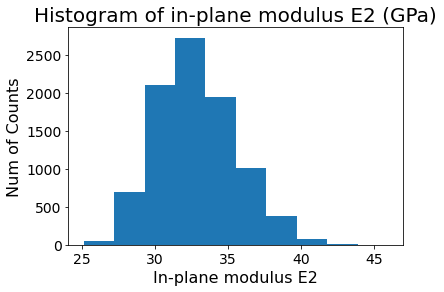}
}
\centering
\subfigure[]{
  \includegraphics[width=0.3\textwidth]{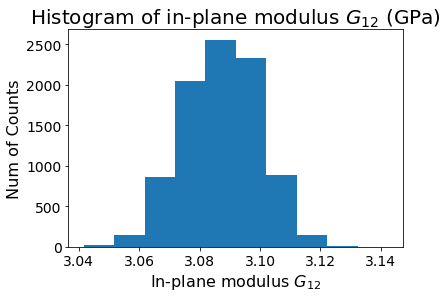}
}
\caption{In-plane modulus distribution for 9000 models of single material woven composite}
\label{img:single_hist}
\end{figure}

\begin{figure}[h!]
\centering
\subfigure[]{
  \includegraphics[width=0.3\textwidth]{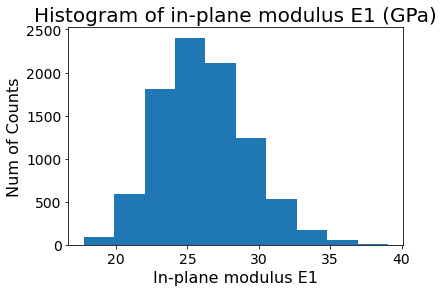}
}
\centering
\subfigure[]{
  \includegraphics[width=0.3\textwidth]{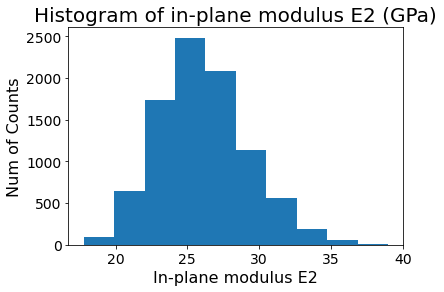}
}
\centering
\subfigure[]{
  \includegraphics[width=0.3\textwidth]{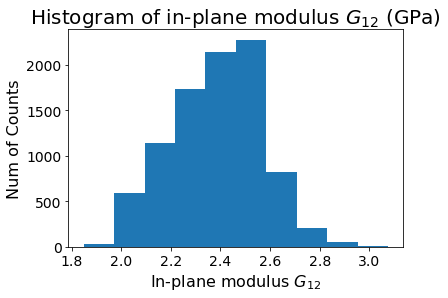}
}
\caption{In-plane modulus distribution for 9000 models of bi-material woven composite}
\label{img:double_hist}
\end{figure}

Different woven composite models could have the same in-plane modulus as discussed earlier. This can be easily visualized through histogram plots, as shown in Figure~\ref{img:single_hist} and Figure~\ref{img:double_hist}. For every single material and bi-material woven composite model, we use 9000 models for verification purposes.

\begin{figure}[h!]
\centering
\subfigure[]{
  \includegraphics[width=0.3\textwidth]{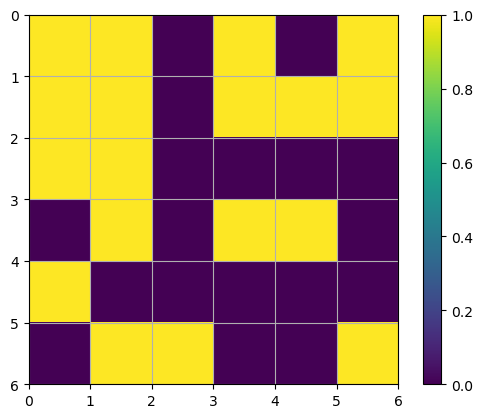}
}
\centering
\subfigure[]{
  \includegraphics[width=0.3\textwidth]{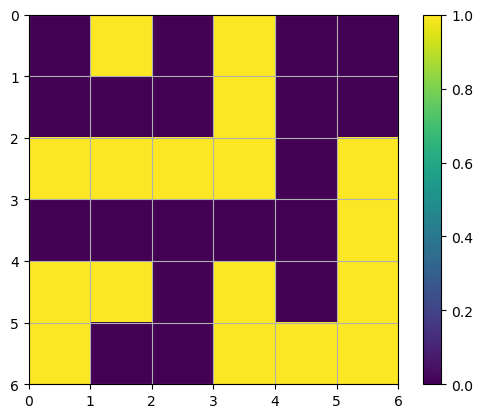}
}
\caption{Single material weave patterns having the same modulus in $E_1$.}
\label{img:same_E1}
\end{figure}

\subsection{Machine Learning Baseline Models}\label{ML_baseline}
\subsubsection{Convolutional-based Encoder-Decoder Network model (Woven-Decoder)} \label{ED_baseline}
\begin{figure}[h!]
\centering
	\includegraphics[width=0.9\textwidth]{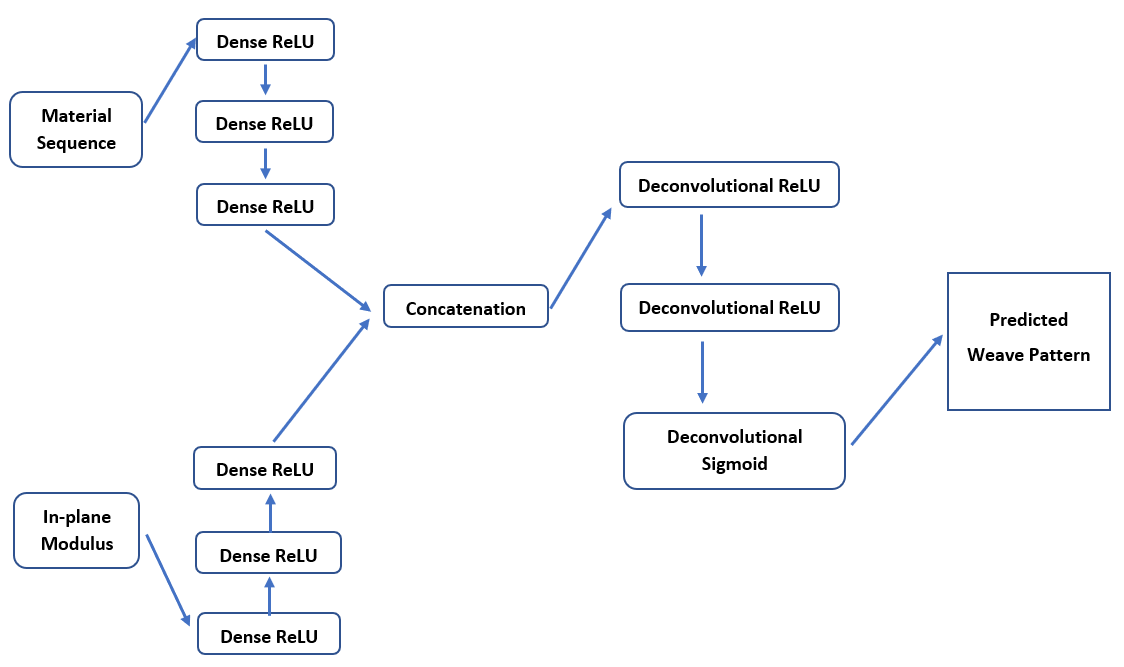}
\caption{Woven-Decoder overall framework for BDPa problem: modules inside green dashed line is the generator and modules inside red dashed line is the discriminator. The 'Deconvolutional' blocks are deconvolutional layers with ReLU or Sigmoid as the activation function and 'Convolutional' blocks are Convolutional layers with ReLU as the activation function.}
\label{img:woven_deconv_ms}
\end{figure}

The Convolutional-based Encoder-Decoder Network is an encoder-decoder neural network that consists of an encoder neural network and a decoder neural network in which one or both are convolutional neural networks. For this paper, the woven-decoder extracts high-level features from two inputs and predicts the results in output physical space. A brief framework overview is shown in Figure~\ref{img:woven_deconv_ms}.

\begin{figure}[h!]
\centering
	\includegraphics[width=0.9\textwidth]{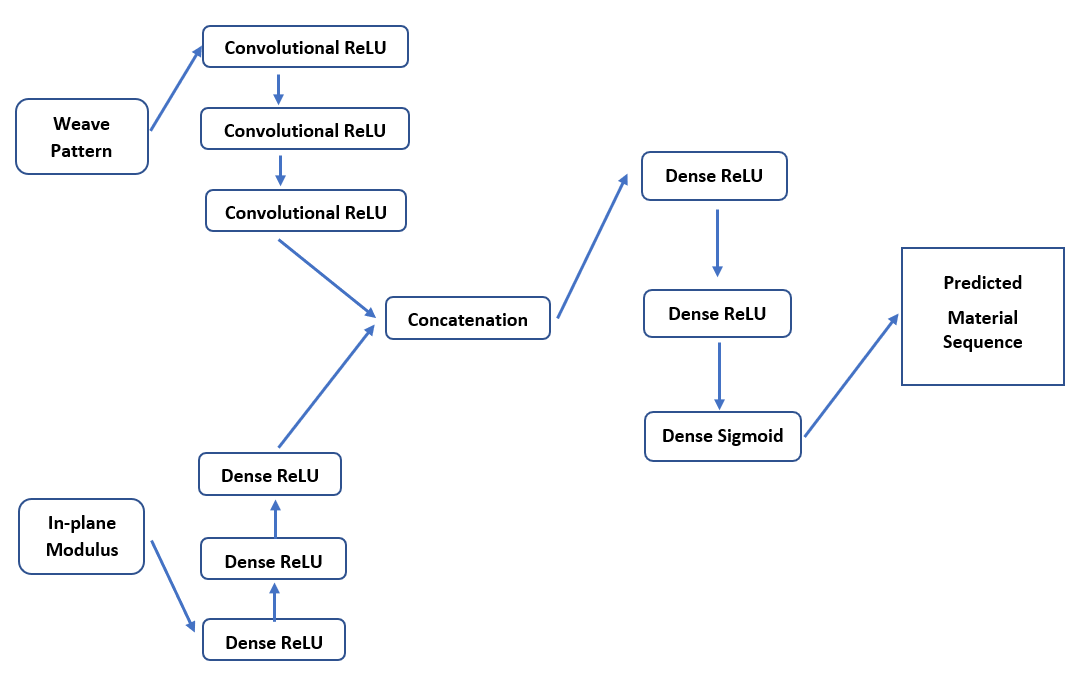}
\caption{Woven-Decoder overall framework for BDPb problem: modules inside green dashed line is the generator and modules inside red dashed line is the discriminator. The 'Convolutional' blocks are Convolutional layers with ReLU as activation function. The Dense layers are connected with either ReLU or Sigmoid activation function.}
\label{img:woven_deconv_ps}
\end{figure}

\subsubsection{Generative Adversarial Network model (Woven-GAN)} \label{GAN_baseline}
\begin{figure}[h!]
\centering
	\includegraphics[width=0.9\textwidth]{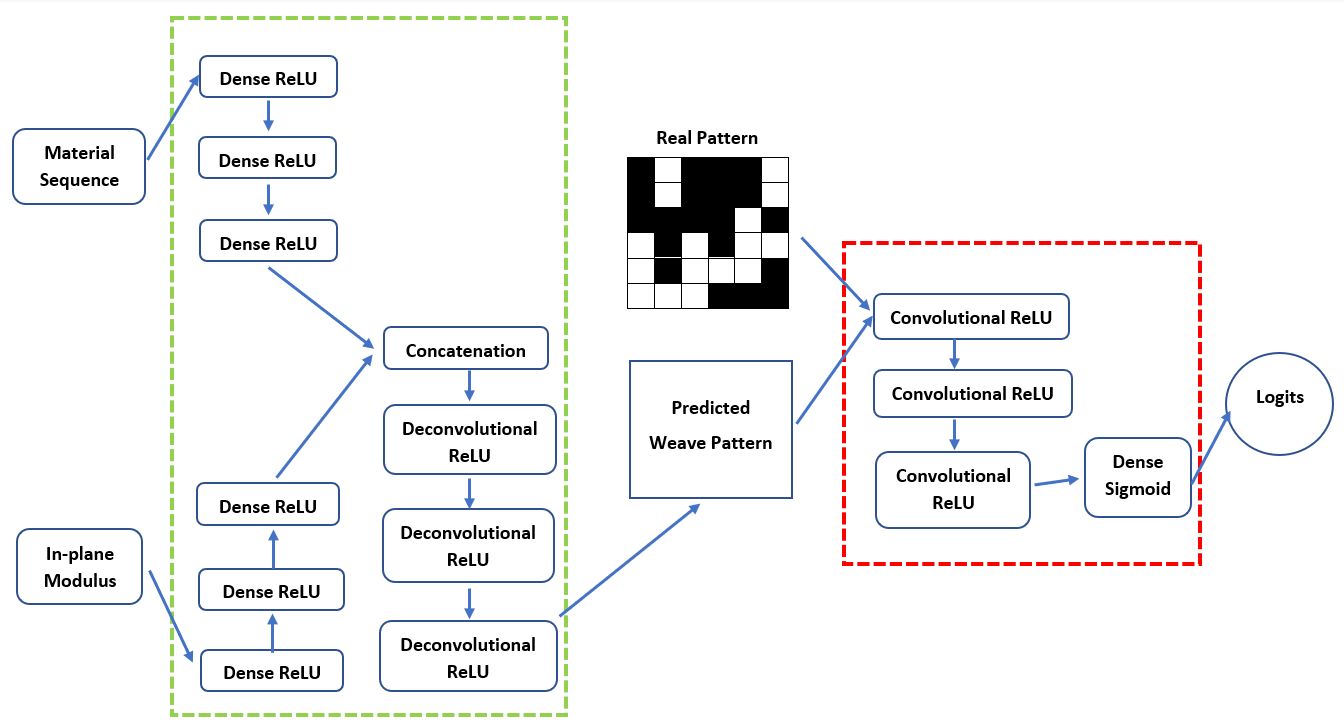}
\caption{Woven-GAN overall framework for BDPa problem: modules inside the green dashed lines are the generator, and modules inside the red dashed line are the discriminator. Each 'Deconvolutional' block consists of Deconvolutional layers with ReLU, and each 'Convolutional' block consists of Convolutional layers with ReLU. The 'logits' module will output the probabilities that our predicted material vector is 'realistic'.}
\label{img:woven_gan_ms}
\end{figure}

\begin{figure}[h!]
\centering
	\includegraphics[width=0.9\textwidth]{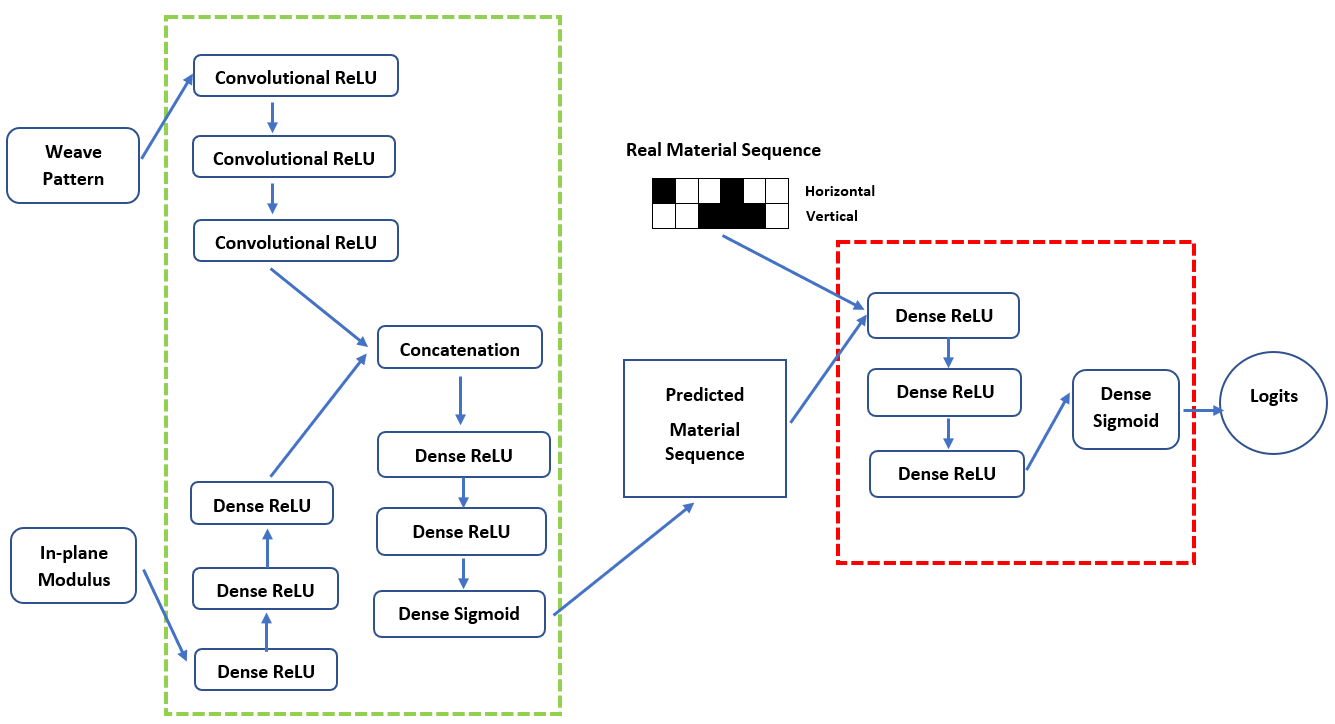}
\caption{Woven-GAN overall framework for BDPb problem: modules inside the green dashed line are the generator, and modules inside the red dashed line are the discriminator. Each 'Convolutional' block consists of Convolutional layers with ReLU. The 'logits' module will output the probabilities that our predicted material vector is 'realistic'.}
\label{img:woven_gan_ps}
\end{figure}

Generative Adversarial Network (GAN) is a class of machine learning framework. GAN consists of a generator and a discriminator. GAN's discriminator tells how much input is realistic, while the generator is used to generate the output that can fool the discriminator. GAN's core idea is 'indirect' training by adding a discriminator model after the prediction, such that the generator can produce a prediction close to the true value. A brief framework overview for the BDPa problem is shown in Figure~\ref{img:woven_gan_ms}, and the framework overview for the BDPb problem is shown in Figure~\ref{img:woven_gan_ps}.

\subsubsection{Genetic Algorithm (Woven-GA)} \label{GA_baseline}
\begin{figure}[h!]
\centering
	\includegraphics[width=0.4\textwidth]{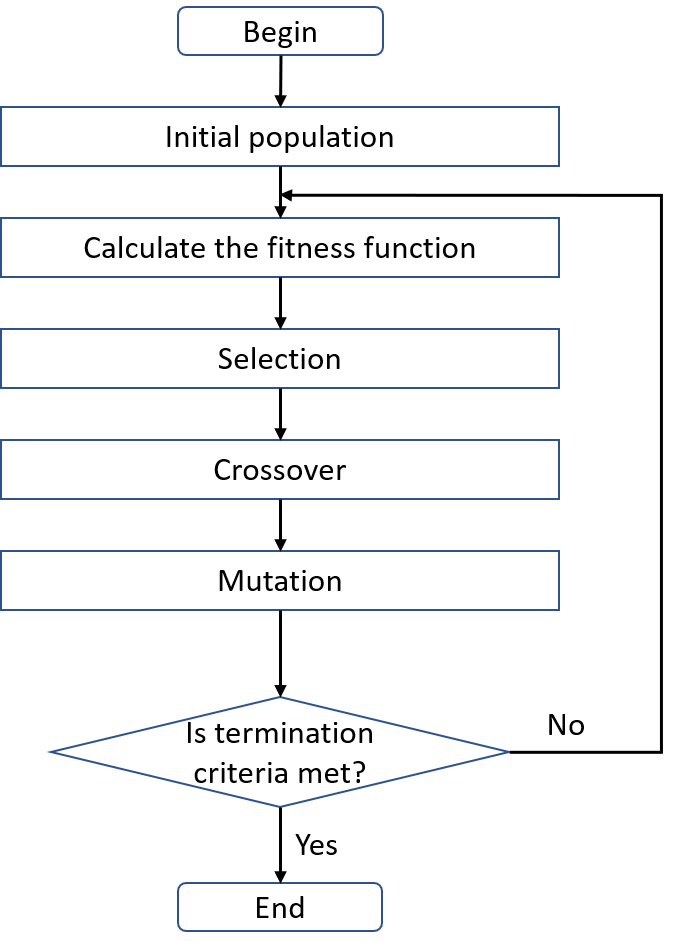}
\caption{Genetic algorithm overall framework.}
\label{img:genetic_algorithm}
\end{figure}

The Genetic Algorithm starts from several randomly generated samples as the first generation and calculates their corresponding values based on the defined objective function, called the fitness function. The crossover module with a predefined rate is performed to determine whether to perform crossover or directly pass the parent into the next generation. Once crossover is performed, the mutation is further used to make the population more diverse to avoid local optima. Such diversity allows the Algorithm to approach global optima faster. This crossover-mutation process will continue as more generations are generated until the termination criteria are met; either we already find the global optima or reach the maximum number of generations. The brief flowchart of the Genetic Algorithm can be summarized in Figure~\ref{img:genetic_algorithm}.

To find out the woven composite architecture with desired in-plane modulus with GA, we define the objective function as Equation~\ref{eqn:ga_target}, where DCNN is the trained Neural Network from the FDP problem. $P_{woven}$, $M_{woven}$ are the woven pattern and material assignment. Here one of them is given, and the other is the prediction from Neural Network. For example, for BDPa, $M_{woven}$ will be given, and $P_{woven}$ will be the predicted woven pattern.

\begin{equation}
    F=||E_{target}-E_{predict}||^2_{2}=||E_{target}-DCNN(P_{woven},M_{woven})||^2_{2}
\label{eqn:ga_target}
\end{equation}

\subsection{Texture feature equations}\label{app:tex-eqn}

This section provides detailed definitions of each statistical feature from GLCM, denoted by matrix $m$.

\begin{equation}
    \mathrm{Contrast}=\sum_{i}\sum_{j}(i-j)^2m(i,j)
\end{equation}

\begin{equation}
    \mathrm{Correlation}=\frac{\sum_{i}\sum_{j}ij[m(i,j)]-\mu_x\mu_y}{\sigma_{x}\sigma_{y}}
\end{equation}

\begin{equation}
    \mathrm{Energy}=\sum_{i}\sum_{j}[m(i,j)]^2
\end{equation}

\begin{equation}
    Homogeneity=\sum_{i}\sum_{j}\frac{m(i,j)}{1+|i-j|}
\end{equation}

\subsection{Proofs of Propositions in Section~\ref{sec:pat_stat_feature}} \label{sec:glcm_proof}

\begin{propositionproof}
We can express the four statistical features of the target GLCM as in Equation~\ref{eqn:glcm_feature}.
\begin{equation}
\begin{aligned}
contrast &= b+c\\
correlation &= \frac{d-\mu_x\mu_y}{\sigma_x\sigma_y}\\
energy &= a^2+b^2+c^2+d^2\\
homogeneity &= a+\frac{b}{2}+\frac{c}{2}+d\\
\end{aligned}
\label{eqn:glcm_feature}
\end{equation}

Where the more generic expression of the statistical features can be found in Appendix~\ref{app:tex-eqn}. In Equation~\ref{eqn:glcm_feature}, there are four independent equations for four elements in GLCM, meaning the four statistical features can uniquely determine the GLCM.
\end{propositionproof}

\begin{propositionproof}
For the same 2-by-2 GLCM as in proof 1, it is evident that $a+b+c+d=C_0$, where $C_0$ is a constant depending on the size of the physical space matrix. Then we can prove that energy tells us the number of transitions in the original space.

We can first consider extreme cases where maximum and minimum energy happens. We can quickly show that maximum energy occurs when one of $(a,b,c,d)$ is non-zero, and all others are zeros, while the minimum energy happens when $a=b=c=d$ (these proofs are shown in Appendix~\ref{app:lemma_proof}). So when we have higher energy, one or two types of transition in physical space must be much larger than the others. Furthermore, when we have small energy, frequencies of different types of transitions will be more evenly balanced.

In addition to our understanding of energy, using other statistical features could further help determine the specific dominated transitions in GLCM. For example, in the largest energy case, a contrast equal to zero tells us a or d is not equal to zero; a correlation smaller than zero tells us $d=0$. Homogeneity square equals to energy tells as $b=c=0$ but $a\, or\, d\neq 0$.
\end{propositionproof}

\begin{propositionproof}
From Proposition~\ref{claim:1}, contrast ($=b+c$) is determined by the sum of off-diagonal terms, while homogeneity ($=a+d+\frac{b+c}{2}$) is determined by the sum of diagonal and discounted off-diagonal terms. Thus, combining contrast and homogeneity will tell us the sum of diagonal ($a+d=$ Homogeneity - Contrast / 2) and off-diagonal terms ($b+c=$ Contrast).
\end{propositionproof}

\subsection{Proofs for Sub-Conclusions used in Proposition~\ref{claim:2} }\label{app:lemma_proof}

In the proofs of Claim 2, we mentioned that GLCM is defined as 
$\begin{bmatrix}
    a & b \\
    c & d
\end{bmatrix}$
. High energy indicates that one or two values in GLCM are much higher than others, and low energy indicates more balanced values. 

\begin{proof}
Finding the highest energy can be described as $argmax _{a,b,c,d} E(a,b,c,d)$, where $E(a,b,c,d)=a^2+b^2+c^2+d^2$, subject to $a+b+c+d=k$. Then rewrite energy expression as $E(a,b,c,d)=a^2+b^2+c^2+(k-a-b-c)^2$. Then by considering the energy as a function of $c$ and by finding $\frac{\partial E}{\partial c}=0$ and noticing $\frac{\partial^2 F}{\partial c^2}=4>0$, the maximum value will happen at $c=k \, or \, 0$. If $c=k$, then $a=b=d=0$, conclusion proved. If $c=0$, we can find $a+b=k$ and thus $d=0$. Then by calculating $\frac{\partial E(a,b)}{\partial a}=0$ and noticing $\frac{\partial^2 F}{\partial a^2}=4>0$, we know $a=0 \, or \, k$. 

Finding the lowest energy can be described as $argmin _{a,b,c,d} E(a,b,c,d)$, where $E(a,b,c,d)=a^2+b^2+c^2+d^2$, subject to $a+b+c+d=C_0$. Like the above proof, we first rewrite energy express as $E(a,b,c,d)=a^2+b^2+c^2+(k-a-b-c)^2$. Then by noticing $\frac{\partial^2 F}{\partial c^2}=4>0$, we want to solve c by $\frac{\partial E}{\partial c}=0$ and find $c=d$. Then we can further rewrite energy expression as $E(a,b)=a^2+b^2+2(\frac{k-a-b}{2})^2$. Then by noticing $\frac{\partial^2 E}{\partial b^2}>0$ we can find $b=c$, thus we must have $a=b=c=d$. 

Conclusion proved.

\end{proof}

\subsection{Example of Modulus Bound Relaxation algorithm}\label{sec:mod_relax}

\begin{figure}[h!]
\centering
\subfigure[]{
  \includegraphics[width=0.3\textwidth]{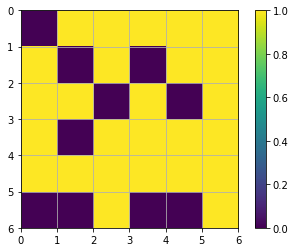}
}
\centering
\subfigure[]{
  \includegraphics[width=0.3\textwidth]{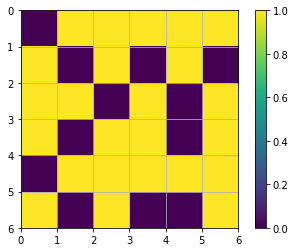}
}
\caption{Weave pattern modification from the Modification Module: (a) original weave pattern predicted from PCNN (b) modified weave pattern from Modification Module}
\label{img:mr_example}
\end{figure}

This section shows how the Modulus Relaxation algorithm modifies the weave pattern predicted for user-defined target moduli values. Figure~\ref{img:mr_example}(a) is the original predicted weave pattern from PCNN, and Figure~\ref{img:mr_example}(b) is the modified weave pattern, which does not have a continuous yarn problem. The original weave pattern was predicted for a target in-plane moduli of $E_1=30GPa, E_2=30GPa, G_{12}=2.5GPa$ with a material vector, $\begin{bmatrix}
    1 & 1 & 1 & 1 & 0 & 0 \\
    0 & 1 & 1 & 0 & 1 & 0
\end{bmatrix}$. The predicted weave pattern has in-plane moduli of $E_1=24.1GPa, E_2=29.49GPa, G_{12}=2.22GPa$. On the other hand, the modified weave pattern having the same material vector has in-plane moduli of $E_1=25.2GPa, E_2=26.8GPa, G_{12}=2.38GPa$, which is close to the modulus of the original predicted weave pattern. The minor reduction in the in-plane moduli for the weave pattern in Figure~\ref{img:mr_example}(b) can be attributed to more undulations between the warp and weft thread. 

\end{document}